\documentclass[preprint,printnumbers,amsmath,amssymb,superscriptaddress,floatfix,aps,prx]{revtex4-1}
\usepackage{graphicx}
\usepackage{colordvi}
\usepackage{color}
\usepackage{xcolor}
\usepackage{bm}
\usepackage[normalem]{ulem}
\usepackage{hyperref}

\begin{document}
\title{Ideal noncrystals: a possible new class of ordered matter without apparent symmetry breaking}

\author{Xinyu Fan}
\affiliation{Department of Physics, University of Science and Technology of China, Hefei 230026, China}
\author{Ding Xu}
\affiliation{Department of Physics, University of Science and Technology of China, Hefei 230026, China}
\author{Jianhua Zhang}
\affiliation{Department of Physics, University of Science and Technology of China, Hefei 230026, China}
\author{Hao Hu}
\affiliation{School of Physics and Optoelectronic Engineering, Anhui University, Hefei 230601, China}
\author{Peng Tan}
\affiliation{State Key Laboratory of Surface Physics and Department of Physics, Fudan University, Shanghai 200433, China}
\author{Ning Xu}
\email{ningxu@ustc.edu.cn}
\affiliation{Department of Physics, University of Science and Technology of China, Hefei 230026, China}
\affiliation{Hefei National Research Center for Physical Sciences at the Microscale and Chinese Academic of Sciences Key Laboratory of Microscale Magnetic Resonance, University of Science and Technology of China, Hefei 230026, China}
\author{Hajime Tanaka}
\email{tanaka@iis.u-tokyo.ac.jp}
\affiliation{Department of Fundamental Engineering, Institute of Industrial Science, University of Tokyo, 4-6-1 Komaba, Meguro-ku, Tokyo 153-8505, Japan
}
\affiliation{Research Center for Advanced Science and Technology, University of Tokyo, 4-6-1 Komaba, Meguro-ku, Tokyo 153-8904, Japan}
\author{Hua Tong}
\email{huatong@ustc.edu.cn}
\affiliation{Department of Physics, University of Science and Technology of China, Hefei 230026, China}

\date{\today}

\clearpage
\begin{abstract}
\bf{Order and disorder constitute two fundamental and opposite themes in condensed matter physics and materials science. Crystals are considered the epitome of order, characterised by long-range translational order. The discovery of quasicrystals, which exhibit rotational symmetries forbidden in crystals and lack periodicity, led to a paradigm shift in solid-state physics. Moving one step forward, it is intriguing to ask whether ordered matter can exist without apparent symmetry breaking. The same question arises considering how ordered amorphous (noncrystalline) solids can be structured. Here, we present the discovery of ideal noncrystals in two dimensions, which are disordered in the conventional sense, lacking Bragg peaks, but exhibit high orderliness based on the steric order, i.e., they are ideally packed. A path-integral-like scheme reveals the underlying long-range structural correlation. We find that these ideal noncrystals are characterised by phononic vibrational modes following the Debye law, fully affine elastic responses, and suppressed density fluctuations at longer wavelengths, reminiscent of hyperuniformity --- all characteristics typically associated with crystals. Therefore, ideal noncrystals represent a peculiar form of matter with a mixed nature---noncrystalline yet possessing crystal-like properties. Notably, these states are found to be thermodynamically favourable, indicating them as a possible new class of ordered matter without apparent symmetry breaking. Our findings significantly broaden the conceptualization of ordered states of matter and may contribute to a deeper understanding of entropy-driven ordering, particularly in generic amorphous materials.}
\end{abstract}

\maketitle
Solids are traditionally classified into two classes based on their structural order, or symmetry. Crystals posses long-range periodic atomic arrangements, leading to a limited number of possible crystalline structures due to the requirement of periodicity, which strongly restricts rotational symmetry (14 Bravais lattices in 3D and 5 in 2D) \cite{ashcroft1976solid}. These simple and ordered structures serves as the foundation for both intuitive understanding and theoretical developments in solid state physics. In contrast, amorphous solids lack specific translational or rotational symmetry, resulting in disordered atomic arrangements~\cite{phillips1981amorphous}. It had been long believed that crystals are the only stable and ordered state of matter at low temperatures, until the discovery of quasicrystals by Shechtman in 1982 \cite{shechtman1984metallic}. Unlike crystals, quasicrystals do not require periodicity and exhibit Bragg peaks in the diffraction pattern with symmetries forbidden by crystallographic laws \cite{levine1984quasicrystals,levine1986quasicrystals,divincenzo1991quasicrystals}. Therefore, the very existence of quasicrystals was initially met with scepticism, and attempts were made to interpret the experimental observations within classical crystallography \cite{pauling1985apparent,steinhardt2019second}. However, the fast accumulation of reports on quasicrystalline phases in metallic compounds, as well as theoretical developments, finally led to a paradigm shift in solid-state physics~\cite{divincenzo1991quasicrystals}. In 1992, the concept of a crystal was refined to include quasicrystals as a second class of well-ordered solids ~\cite{international1992report}. It is now known that quasicrystals are not rare but rather common in various systems, including atomic alloys \cite{divincenzo1991quasicrystals}, soft matter \cite{zeng2004supramolecular,iacovella2011self}, and granular materials \cite{plati2024quasi}. Actually, quasicrystals were prepared unknowingly long before Shechtman's discovery~\cite{steurer2018quasicrystals}, partially due to the limitation of X-ray diffraction techniques, but more importantly, the unawareness of such a possibility. This underscores the caution that other forms of order might pass unnoticed due to the lack of lens and angle to detect them. Given the limitation of diffraction techniques widely used for structure characterizations~\cite{steurer2018quasicrystals}, an intriguing question arises: Does ordered matter exist without apparent breaking of both translational and rotational symmetries?

The quest for order other than long-range periodicity is also a central theme in the study of supercooled liquids and amorphous solids, i.e., glasses~\cite{steinhardt1983bond,alexander1998amorphous,tanaka2012bond,tanaka2019revealing}. Interestingly, the early theoretical developments of quasicrystals were actually entangled with this quest~\cite{steinhardt1981icosahedral,steinhardt1983bond}. Motivated by the observation of extended icosahedral order in supercooled liquids, the possible existence of long-range bond-orientational (i.e., rotational) order approaching the glass transition was explored \cite{steinhardt1981icosahedral,steinhardt1983bond,nelson1983order}. Also inspired by the Penrose tilling~\cite{penrose1974role}, Levine and Steinhardt coined the concept of quasicrystals and formulated the first theory for this new class of ordered matter~\cite{levine1984quasicrystals,levine1986quasicrystals}. Notably, this conceptual breakthrough was achieved independent of experimental evidence. However, since quasicrystals typically form in binary or ternary compounds with specific relations of particle sizes through a first-order phase transition~\cite{shechtman1984metallic}, they are closer to crystals rather than the `amorphous order' quested in generic glass formers~\cite{berthier2011theoretical}. This raises fundamental questions about the extent of order achievable in amorphous (noncrystalline) solids and the nature of order that may exist when both translational and rotational symmetries are lacking.

Thermodynamics tells us that structure ordering is ultimately driven by free energy. In systems where structure arises from steric constraints on particle packing~\cite{frank1958complex,bernal1959geometrical,torquato2018perspective}, free energy is mainly controlled by entropy \cite{tanaka2019revealing}. Sterically favoured structures provide more room for particle vibration, thereby maximising vibrational entropy~\cite{tanaka2012bond}. One illustrative example of this entropy-driven ordering phenomenon is the crystallization of hard spheres. Many glass formers with simple isotropic interactions, including colloids, granular materials, and metals~\cite{frank1958complex,bernal1959geometrical,torquato2018perspective,xing2024origin} fall into this category~\cite{tanaka2019revealing}. Recent studies have shown that in simple glass formers with strong frustrations against crystalline or quasicrystalline order, subtle sterically favoured structures, identified by a new structural order parameter characterising local packing capability, generally grow with decreasing temperature~\cite{tong2018revealing,tong2019structural}.
These observations underscore the thermodynamic favourability of steric ordering.
However, the quality of such steric order is limited by intrinsic frustrations in the model glass systems and the drastic kinetic slowing-down at low temperatures. This precludes a clear identification of the possible existence of ideal steric order (other than a crystal or quasicrystal) and its associated physical properties. Fundamentally, the primary obstacle in this exploration lies in the lack of understanding of how to construct an ideal sterically favoured structure.

\begin{figure*} 
 \begin{center}
    \includegraphics[width=0.98\textwidth]{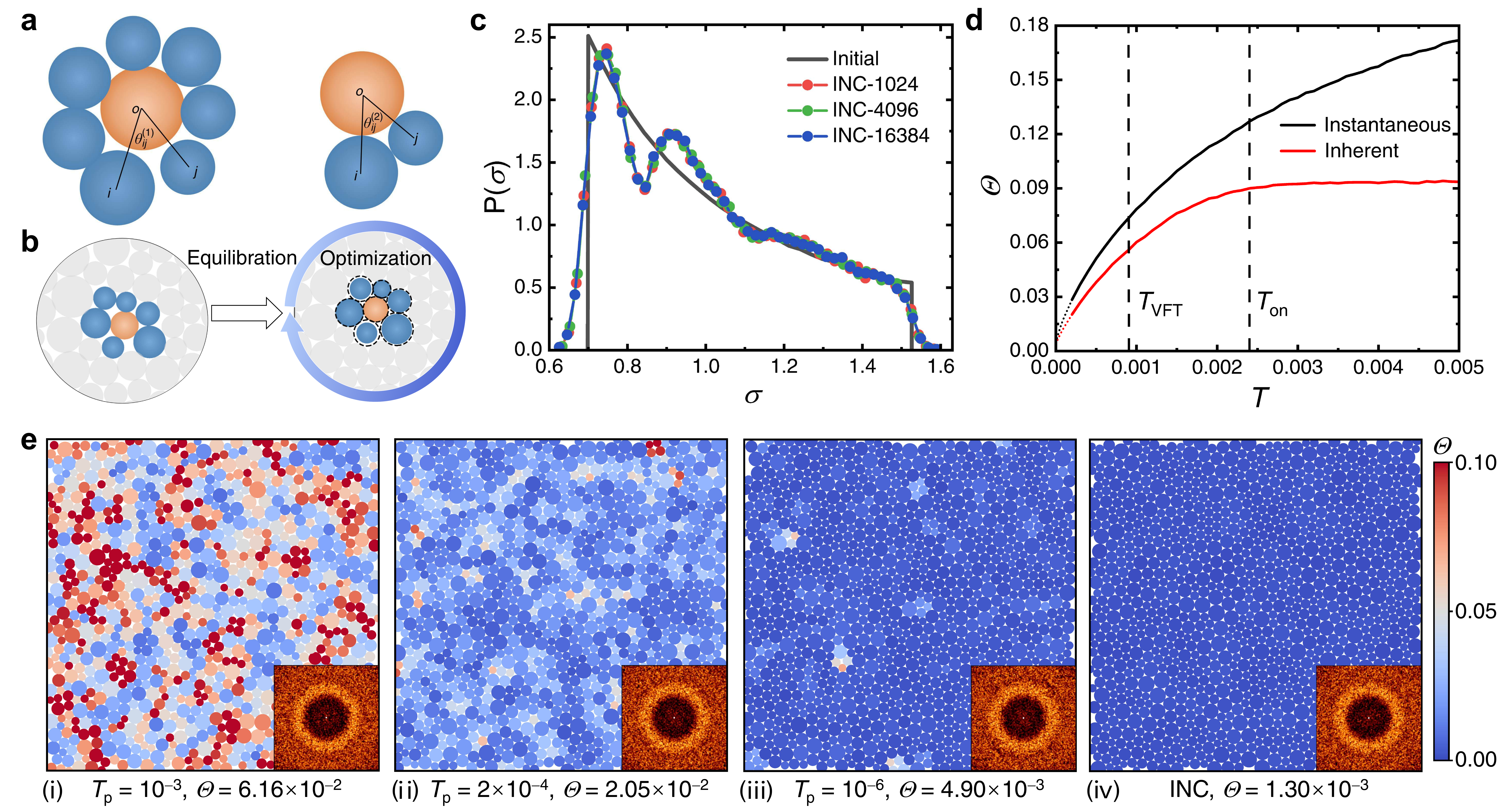}
    \caption{\textbf{Realization of ideal noncrystals with optimal steric order.} \textbf{a,} Definition of the structural order parameter $\Theta$, characterising the capability of efficient packing (Methods)~\cite{tong2018revealing}. A typical local particle configuration is shown along with a reference configuration where the three highlighted particles are just touching each other. \textbf{b,} Schematic of the procedures to achieve optimal steric order. The system is first equilibrated in the liquid state and then slowly cooled to zero temperature using the swap Monte Carlo algorithm (SMC; Methods). Then, the particle sizes are adjusted to minimise $\Theta$ (dashed circles), followed by relaxation to mechanical equilibrium. The optimization procedure is repeated until $\Theta$ no longer decreases (Methods).
    \textbf{c,} Particle diameter distributions before (a truncated power-law distribution $P(\sigma)\sim \sigma^{-2}$; Methods) and after the optimization procedure (INC, indicating ideal noncrystals). The collapse of data from different system sizes suggests the absence of finite-size effects. Ideal noncrystals are then melted and cooled down using SMC with a cooling rate $dT/dt = 10^{-10}$. \textbf{d,} Evolution of steric order $\Theta$ in instantaneous and inherent states upon cooling. The onset temperature ($T_{\rm on}=2.4\times 10^{-3}$) and glass transition temperature according to the Vogel-Fulcher-Tammann law ($T_{\rm VFT}=9.1\times 10^{-4}$) are indicated by vertical dash lines as references (Extended Data Fig.~2). The system falls out of equilibrium at $T=2\times10^{-4}$ (Extended Data Fig.~3), below which data are shown by dotted lines. \textbf{e,} Visualisation of the steric order $\Theta$ of inherent-state configurations from different parent temperatures $T_p$. Insets: Two-dimensional static structure factor indicating no crystalline order.}
    \label{fig1}
 \end{center}
\end{figure*}

Our approach to addressing the above questions is to parameterise frustrations against ideal steric order in a manner that allows for optimization towards zero~\cite{nelson1983order}. Central to this approach is our definition of a structural order parameter, denoted as $\Theta$, which quantifies the capability of efficient packing, regardless of whether it is associated with crystalline order featuring discrete symmetry breaking or more subtle `amorphous order' lacking apparent symmetry breaking (Fig.\ref{fig1}a and Methods)~\cite{tong2018revealing,tong2019structural}.
Another essential aspect of our approach is our focus on thermodynamically favourable structures, which is crucial for structuring in natural materials~\cite{tanaka2012bond}. To this end, we begin with a two-dimensional model glass former with a power-law particle size distribution to prevent both crystallization and phase separation (Fig.~\ref{fig1}b,c and Methods)~\cite{ninarello2017models}.
This setup allows for sufficient equilibration using the swap Monte Carlo algorithm (SMC) down to very low temperatures~\cite{ninarello2017models}, resulting in a system with a high degree of steric order. Subsequently, we optimise the particle sizes to achieve ideal steric order (Fig.~\ref{fig1}b, Extended Data Fig.~1, and Methods). This approach minimises disturbance to the pre-equilibrated system, ensuring that the obtained states are at, or at very close to, thermodynamic equilibrium.
The final particle size distribution is close to the initial one (Fig.~\ref{fig1}c). The polydispersity changes slightly from $\Delta=23.4\%$ to $23.2\%$.
The size ratio between smallest and biggest particles decreases only a little from $\sigma_{\rm min}/\sigma_{\rm max}=0.448$ to 0.423, which is larger than the value expected if the small particle was positioned within the interstices among big particles (e.g., an interstitial particle surrounded by four bigger ones would yield $\sigma_{\rm min}/\sigma_{\rm max}\simeq  0.414$). This is a crucial requirement from Gibbs's definition of a pure phase~\cite{gibbs1878equilibrium}.
It is worth noting that these fundamental aspects set our work apart from previous studies where the particle size was treated as an additional degree of freedom to optimise the structure~ \cite{brito2018theory,kim2022structural,kim2024dense,corwin2024ideal}.
Consequently, we term the resulting states `ideal noncrystals' (INC), as their subsequent analysis reveals a remarkable combination of sterically perfect yet noncrystalline structure, accompanied by properties reminiscent of crystals.

To confirm that the ideal-noncrystal states are indeed thermodynamically favourable, we subject the configurations obtained from optimization procedures to melting and then follow the evolution of the resulting liquid state upon slow cooling. The steric order parameter $\Theta$, both for instantaneous and inherent states, decreases with temperature towards an extremely low value, indicating the approach to ideal-noncrystal states (Fig.~\ref{fig1}d). The development of steric order is further illustrated by the spatial distribution of $\Theta$ (Fig.~\ref{fig1}e). The configurations obtained at low temperatures by slow cooling (Fig.~\ref{fig1}e(iii)) can be regarded as an ideal noncrystal (Fig.~\ref{fig1}e(iv)) with several localised defects. Such defects are inevitable in practice, akin to vacancies or dislocations in crystals. The absence of crystalline order is confirmed by the two-dimensional static structure factor (insets of Fig.~\ref{fig1}e). Notably, the unjamming volume fraction of ideal noncrystals is $\phi\simeq 0.909$, higher than that of the hexagonal close packing $\phi_{\rm cp}\simeq 0.907$. Therefore, size fractionation and crystallization are intrinsically unfavourable in our system (Methods and Supplementary Information).

\begin{figure*}
  \begin{center}
    \includegraphics[width=0.75\textwidth]{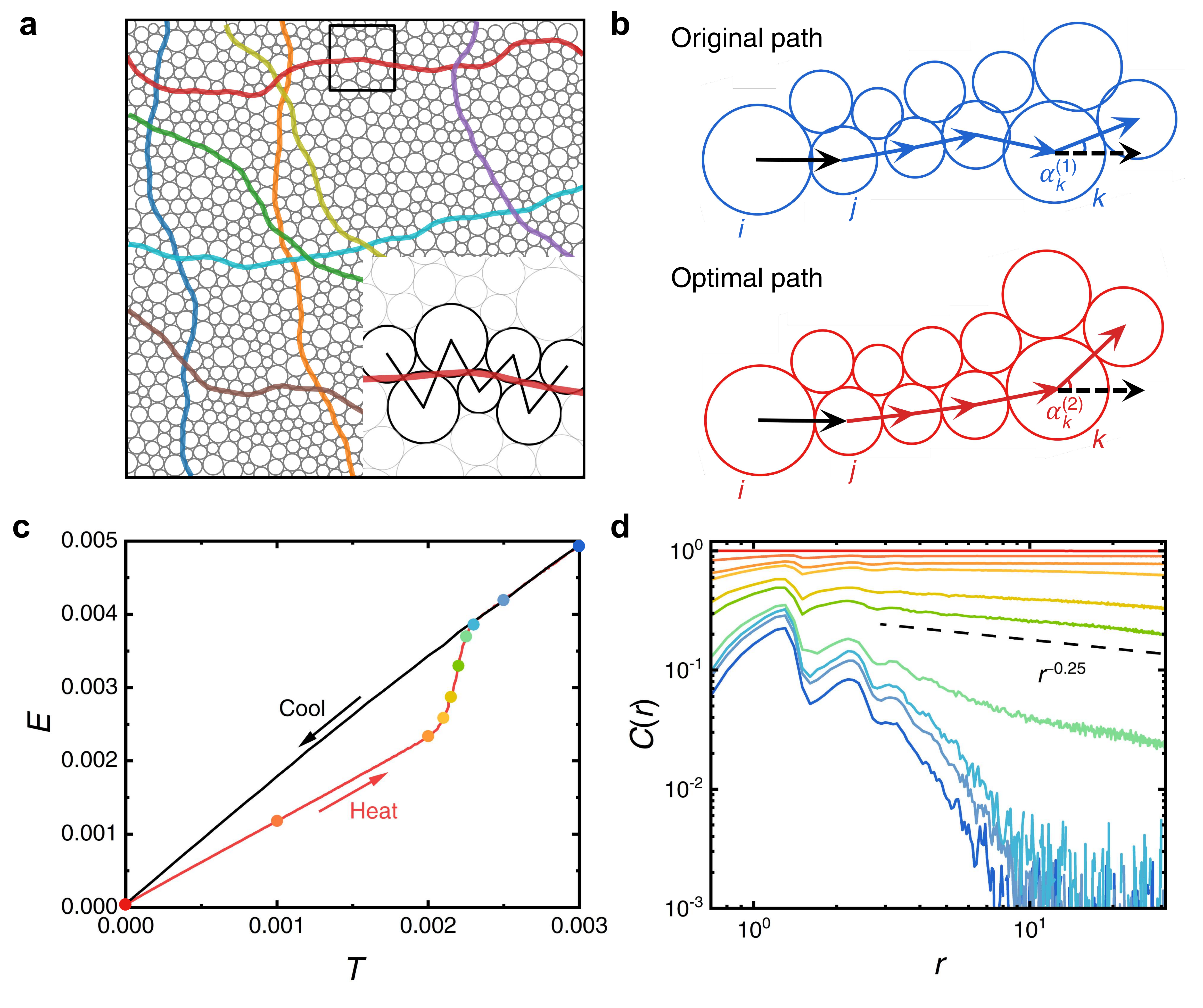}
    \vspace{-6mm}
    \caption{\textbf{Long-range structural correlation in ideal-noncrystal states.} \textbf{a,} Selected examples of coherent paths in an ideal noncrystal configuration, which can extend and densely cover the whole system. Inset: Definition of the coherent path. Based on Voronoi-Delaunay tessellation, the zigzag patterns of triangles formed by three neighbouring particles can be identified. The coherent path goes through these zigzag patterns, crossing intersections of Voronoi and Delaunay tessellations. \textbf{b,} Definition of the path-integral-like correlation function. An optimal path is constructed from the original coherent path by arranging particles into perfectly touching triangle units. Arrows connect particles on one side of the coherent path, with the initial pair of particles marked as $i$ and $j$, and another particle in the sequence as $k$. The direction of ${\bf r}_{ij}$ (black solid and dashed arrows in both the original and optimal path) serves as the reference, and the winding angle of the arrow from particle $k$ to its subsequent neighbour is defined as the accumulated rotational angle along the path. The difference between winding angles in the original and optimal paths measures structural decorrelation along the coherent path (Methods). \textbf{c,} Evolution of potential energy per particle $E$ when ideal-noncrystal configurations are heated using normal MD at a rate of $dT/dt=10^{-10}$, followed by cooling using SMC at a rate of $dT/dt=10^{-10}$. \textbf{d,} The path-integral-like correlation function $C(r)$ (Methods) for the state points indicated in \textbf{c}. The dashed line represents $C(r)\sim r^{-0.25}$, predicted by the KTHNY theory of 2D melting from the hexatic phase to liquids \cite{strandburg1988two}.}
    \label{fig2}
  \end{center}
\end{figure*}

Clearly, ideal noncrystals exhibit no apparent breaking of both translational and rotational symmetries (Fig.~\ref{fig1}e(iv) and the inset). Therefore, there is no established method to characterise the underlying structural correlation. Inspired by the remarkable coherence observed along the contour of particles (Fig.~\ref{fig2}a), we have developed a new scheme to extract this nonconventional structural correlation.

First, we employ the Voronoi-Delaunay tessellation to formulate a rigorous definition of `coherent paths' (Fig.~\ref{fig2}a and Methods). These coherent paths are akin to lattice planes in crystals. In essence, the idea is that every three neighbouring particles form a fundamental structural unit---a triangle---in 2D. These triangles cooperate under steric constraints, propagate throughout space, and mediate interactions. We thus introduce a path-integral-like correlation function to quantify the structural correlation along the coherent path, by tracking the decoherence of triangles (Fig.~\ref{fig2}b and Methods). We study the evolution of ideal noncrystals when heated at a constant heating rate using molecular dynamics simulations (MD) and then cooled down using SMC (Fig.~\ref{fig2}c). The steep increase in potential energy per particle $E$ during the melting of ideal noncrystals highly resembles that of monodisperse crystals, suggesting the ultrastability of ideal noncrystals, whereas the freezing process does not (Extended Data Fig.~5a). This is an interesting observation which deserves future in-depth investigations. Here, to be concise, we focus on structural evolution during the melting process. Remarkably, long-range structure correlations are observed before melting, which turn to be short-range exponential decay in liquid states (Fig.~\ref{fig2}d). At intermediate temperatures, the correlation appears to have a power-law form $C(r)\sim r^{-\eta}$, with the exponent decreasing to around $\eta=0.25$ just before entering the liquid phase~\cite{halperin1978theory}.

We have also applied $C(r)$ to characterizing the two-dimensional melting of monodisperse crystals and found that it gives essentially the same information as the spatial correlation of hexatic bond-orientational order parameter $\Psi_6$ (Extended Data Fig.~5b-d)~\cite{strandburg1988two}. This validates the efficacy of our methodology.
More importantly, these observations point to an intimate similarity between ideal noncrystals and hexatic crystals and strongly support ideal noncrystals as a new class of ordered matter without apparent symmetry breaking.
The question of whether a continuous phase transition (akin to the Kosterlitz-Thouless transition \cite{kosterlitz1973ordering} but of a nonequilibrium nature as suggested by Extended Data Fig.~6), or the Kosterlitz-Thouless-Halperin-Nelson-Young (KTHNY) scenario of two-step transition~\cite{kosterlitz1973ordering,halperin1978theory,nelson1979dislocation,young1979melting}, describes the melting of ideal noncrystals remains an intriguing problem for future study.

\begin{figure*}
  \begin{center}
    \includegraphics[width=0.95\textwidth]{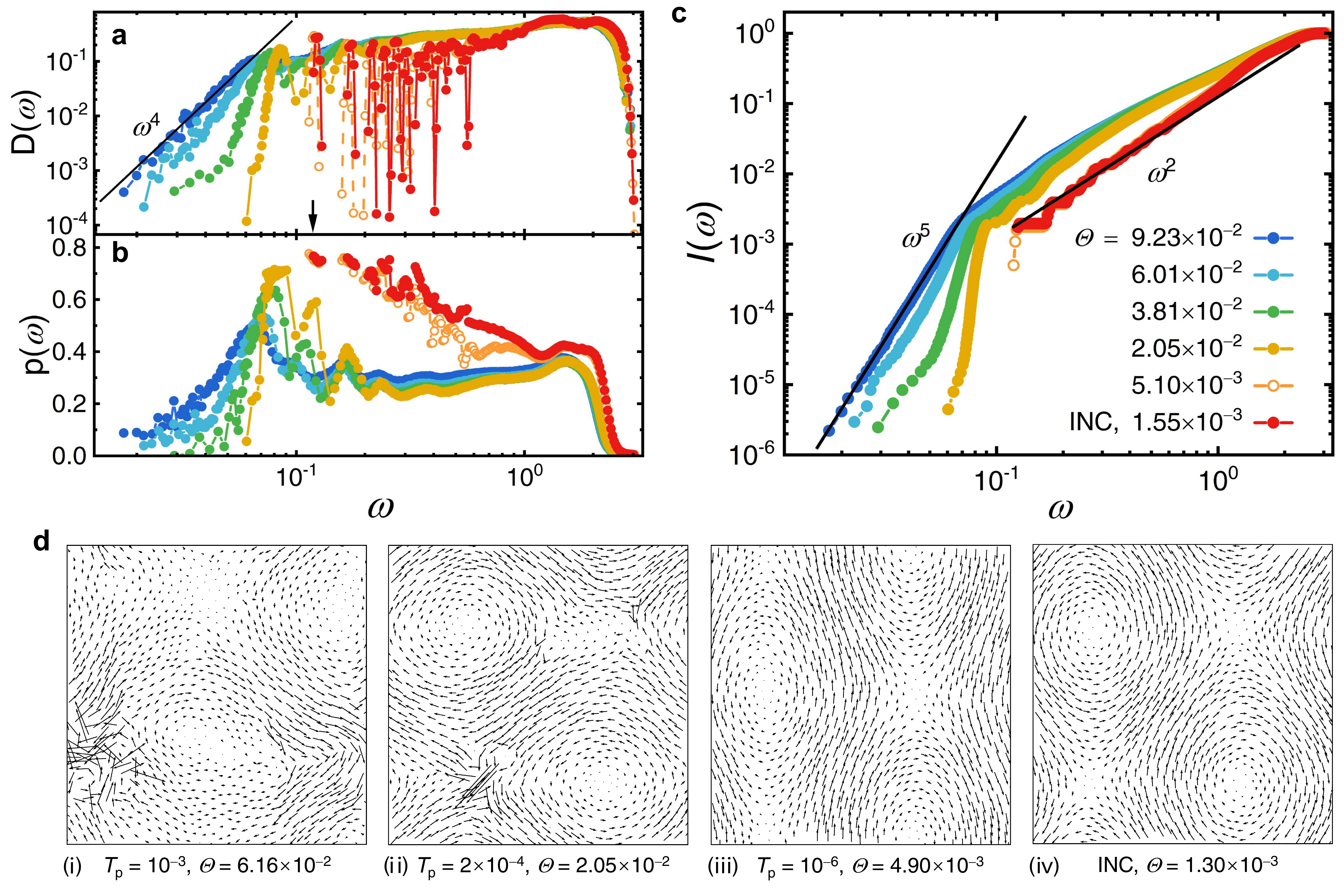}
    \caption{\textbf{Evolution of vibrational modes approaching ideal noncrystals.} \textbf{a}-\textbf{c,} Vibrational density of states $D(\omega)$ (\textbf{a}), and the corresponding participation ratio $p(\omega)$ (\textbf{b}) and integrated vibrational density of states $I(\omega)$ (\textbf{c}) for systems with different degrees of steric order $\Theta$ (Methods). The arrow in \textbf{a} denotes the frequency of the first plane-wave mode in ideal noncrystals according to the Debye theory, and the solid line indicates $D(\omega)\sim \omega^4$ typical for amorphous solids. The two solid lines in \textbf{c} indicate the typical behaviours of amorphous solids $I(\omega)\sim\omega^5$ and crystals $I(\omega)\sim\omega^2$, respectively. \textbf{d,} Corresponding to Fig.~\ref{fig1}e, visualization of the lowest-frequency modes for configurations with different degrees of steric order. The vibrational modes approach phonons in crystals when approaching the ideal-noncrystal states.}
    \label{fig3}
  \end{center}
\end{figure*}

The distinct structure of ideal noncrystals manifests in distinct physical properties. We first investigate the vibrational states, a fundamental characteristic of solids linked to various low-temperature properties (Fig.~\ref{fig3}, Methods)~\cite{ashcroft1976solid,phillips1981amorphous}. As steric order develops towards ideal noncrystals, the low-frequency vibrational density of states transitions from $D(\omega)\sim \omega^4$ (Fig.~\ref{fig3}a), typical for amorphous solids contributed by the non-phononic quasilocalised modes (Fig.~\ref{fig3}b,d(i)) \cite{lerner2016statistics,richard2020universality}, towards Debye scaling $D(\omega)\sim \omega^{d-1}$ (here $d=2$ for 2D, as deduced from the integrated vibrational density of states shown in Fig.~\ref{fig3}c), typical for crystals where the vibrational states are phonons \cite{ashcroft1976solid}. We find that the frequencies of the lowest-frequency modes in ideal noncrystals follow the Debye prediction (indicated by the arrow in Fig.~\ref{fig3}a, Methods), with a high participation ratio (Fig.~\ref{fig3}b), and real-space visualizations confirm their plane-wave feature (Fig.~\ref{fig3}d(iii),(iv)). Therefore, the vibrational modes in ideal noncrystals resemble phonons in crystals. While phonons can be understood as Goldstone modes associated with the breaking of translational and rotational symmetries in crystals~\cite{ashcroft1976solid}, it is not immediately clear whether the phonon-like modes in ideal noncrystals are linked to the breaking of a subtle symmetry. One possible candidate is the long-range correlated structure along the coherent paths (Fig.~\ref{fig2}), but this requires more elaborated theoretical efforts to validate. We note that the vibrational modes in configurations obtained by slow cooling ($\Theta\approx 5.1\times 10^{-3}$ in Fig.~\ref{fig3}) are close to those in ideal noncrystals, suggesting that the nature of ideal noncrystals remains robust against certain degrees of imperfectness.

\begin{figure*} 
  \begin{center}
    \includegraphics[width=0.9\textwidth]{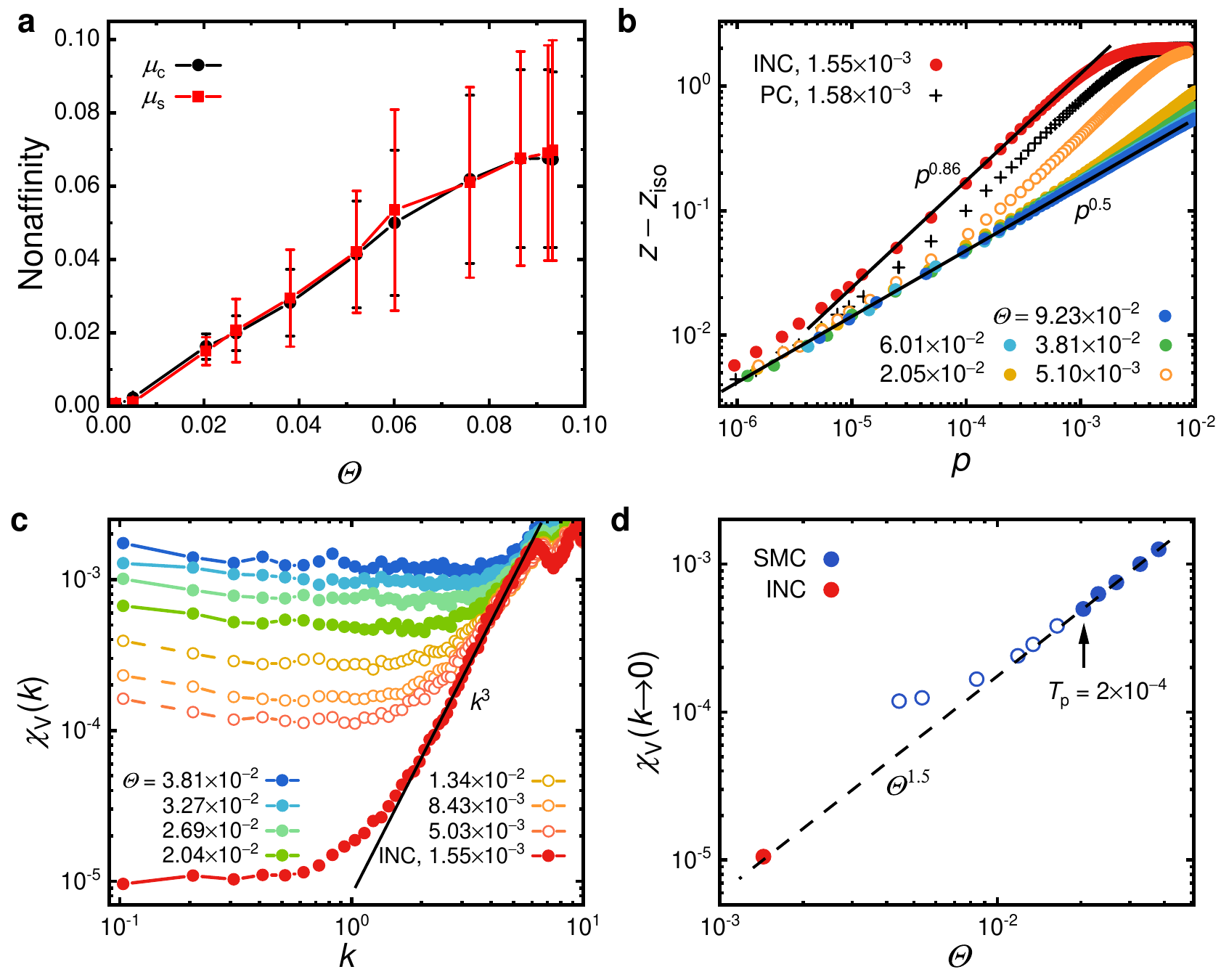}
     \caption{\textbf{Evolution of elasticity and hyperuniformity approaching ideal noncrystals.} \textbf{a,} Nonaffinity as a function of steric order $\Theta$ under isotropic compression ($\mu_{c}$) or simple shear ($\mu_{s}$; Methods). \textbf{b,} Excess coordination number $z-z_{\rm iso}$ as a function of pressure $p$ for systems with different degrees of steric order $\Theta$. Results from crystals with extremely weak polydispersity $\Delta=0.231\%$ (PC; Extended Data Fig.~7), having approximately the same $\Theta$ as ideal noncrystals (INC), are shown as a reference. The comparison suggests that our ideal noncrystals are more ordered than these polydisperse-crystal states. The two solid lines indicate typical scaling behaviour of amorphous solids when approaching the jamming transition $z-z_{\rm iso}\sim p^{0.5}$, and that previously observed in weakly polydisperse crystals when decompressed from high pressures $z-z_{\rm iso}\sim p^{0.86}$ \cite{tong2015crystals}, respectively. \textbf{c,} Spectral density $\chi_V(k)$ for systems with different degrees of steric order $\Theta$ (Methods). The solid line indicates ideal hyperuniformity $\chi_V(k)\sim k^3$. \textbf{d,} The plateau value of $\chi_V(k)$ when approaching the low-$k$ limit as a function of steric order $\Theta$. The arrow indicates the lowest temperature that can be equilibrated by SMC. The black dashed line represents a power-law extrapolation of equilibrium data points $\chi_V(k\rightarrow 0)\sim \Theta^{1.5}$, which passes through that of ideal-noncrystal states.
Out-of-equilibrium data sets are shown by open symbols in \textbf{b}-\textbf{d}.}
    \label{fig4}
  \end{center}
\end{figure*}

Elasticity represents another fundamental characteristic of solids. In crystals, the elastic response to simple shear or compression is trivially affine. In contrast, in amorphous solids, it is highly nonaffine, resulting in a number of peculiar elastic and plastic behaviours \cite{maloney2006amorphous,van2009jamming}. Here, we quantify the nonaffinity by accessing the relative strength of nonaffine to affine displacements (Methods) \cite{tong2015crystals}. We find that, with the decrease of steric order parameter $\Theta$, nonaffinity decreases to nearly zero in ideal noncrystals (Fig.~\ref{fig4}a). Since nonaffinity can be considered a consequence of structural disorder, this result suggests that ideal noncrystals represent an exotic ordered state of matter, akin to crystals in terms of their mechanical aspect.

Jamming provides a paradigm to understand how rigidity emerges in amorphous solids, accompanied by unusual properties distinct from crystals~\cite{van2009jamming,liu2010jamming}. Recently, by studying crystals with extremely weak polydispersity (Extended Data Fig.~7), we discovered new scaling behaviours on the edge of perfect crystalline order \cite{tong2015crystals}. Here, we characterise the evolution of excess coordination number $z-z_{\rm iso}$ ($z_{\rm iso}=2d-2d/N$ is the isostatic value according to the Maxwell criterion for rigidity) with decreasing pressure $p$ and investigate how its scaling relation with pressure depends on the steric order (Fig.~\ref{fig4}b). When approaching ideal noncrystals, we observe a new scaling relation $z-z_{\rm iso}\sim p^{0.86}$, different from the typical jamming scaling $z-z_{\rm iso}\sim p^{0.5}$, akin to weakly polydisperse crystals \cite{tong2015crystals}.
This result again suggests that ideal noncrystals are as ordered as crystals, although in different manners.


Let us revisit the structure and consider a particular form of structural order related to the suppression of large-scale density fluctuations, i.e., hyperuniformity \cite{torquato2018hyperuniform,torquato2018perspective}.
Besides perfect crystals and quasicrystals, hyperuniformity has been observed in exotic disordered materials that are usually far from thermodynamic equilibrium~\cite{torquato2018hyperuniform}, such as the maximally random jammed packings~\cite{zachary2011hyperuniform} and active matter systems~\cite{lei2019nonequilibrium,chen2024emergent}. Hyperuniformity can also emerge in systems with delicate long-range or many-body interactions aimed at achieving disordered ground states \cite{torquato2015ensemble}.
Because the development of steric order ensures better packing efficiency, it is natural to expect the suppression of density fluctuations and the emergence of hyperuniformity in our systems. Here, we characterise the spectral density $\chi_V(k)$ to adequately capture the fluctuations in local volume fraction in polydisperse configurations~\cite{torquato2018hyperuniform}. An asymptotic power-law scaling relation $\chi_V(k)\sim k^3$ is observed in the low $k$ regime as we approach the ideal-noncrystal states (Fig.~\ref{fig4}c), reminiscent of class I hyperuniformity~\cite{torquato2018hyperuniform}. Due to the inevitable presence of imperfections, ideal hyperuniformity is not exactly reached, and a plateau appears as $k\rightarrow 0$, deviating from the power-law relation.
We measure the plateau value of $\chi_V(k)$ (averaged over the five lowest-$k$ points) and examine its relation to $\Theta$ (Fig.~\ref{fig4}d). Interestingly, a power-law relation $\chi_V(k\rightarrow 0)\sim \Theta^{1.5}$ is found to describe the equilibrium data points from slowing cooling (solid blue circles) as well as that from an ideal noncrystal (solid red circle). Its value may be interpreted as the amount of defects in terms of local density. The same behaviour is also observed in crystals with the weak disorder (Extended Data Fig.~7)~\cite{kim2018effect}. This result implies the equilibrium nature of the obtained ideal-noncrystal states and suggests an underlying hyperuniformity as $\Theta$ goes to zero.
Nevertheless, we do not observe a power-law regime for $k<1$. Further careful investigation is necessary to verify our conjecture.

In summary, we have demonstrated the existence of a unique class of ordered matter without apparent symmetry breaking.
Conceptually, this represents a natural extension of the notions of crystals and quasicrystals, exhibiting perfect steric order with a particular type of long-range correlations yet noncrystalline structures. We show that their vibrational modes are phononic following the Debye law, their elastic responses are fully affine, they suppress large-scale density fluctuations approaching hyperuniformity, and they are thermodynamically favourable. Therefore, we term these states `ideal noncrystals', underlining the mixed nature of noncrystalline structure and crystal-like properties. This finding is expected to constitute an important advancement towards a more comprehensive understanding of ordered states of matter.
Considering that ideal noncrystals can be realized through thermodynamic processes and are robust against certain imperfections, their realization in modern colloidal experiments~\cite{yunker2014physics} or through 3D printing techniques~\cite{corker2019} appears feasible. Actually, pieces of ideal noncrystals mixed with crystals or amorphous solids might have been prepared unknowingly in previously experimental or numerical studies. Our methodology may provide the lens to identify them in future studies. Although the existence of ideal noncrystals in three dimensions remains to be explored, our preliminary result that a high concentration of distorted icosahedra exists in a highly polydisperse glass former suggests its feasibility. While our study has focused on the basic structural, elastic, and vibrational properties of ideal noncrystals, an interesting avenue for future research lies in exploring their full range of properties, including their stability, as well as their thermal, phononic, and photonic properties, which may hold significant practical importance.

The achievement of ideal noncrystals, driven by entropy through the optimization of a generic steric order parameter, is a thermodynamic necessity in systems like hard spheres. This phenomenon is not limited to simple model glass formers but also extends to real-world materials such as colloids, granules, and notably, atomic systems like metallic alloys \cite{frank1958complex,bernal1959geometrical,torquato2018perspective}. Therefore, in addition to crystalline or quasicrystalline order, the propensity for noncrystalline order---despite its apparent disorder---should be considered with equal importance when examining structure ordering in supercooled liquids and glasses.
An immediate question arises concerning the relationship between ideal noncrystals and ideal glasses, a central concept in the physics of glass transition~\cite{berthier2011theoretical}. It is important to note that the concept of ideal glasses is primarily defined within mean-field theories, where structure ordering is not considered. This concept, originating from random-first-order phase transitions, is hypothesised to apply generally to fragile liquids~\cite{berthier2011theoretical}.
In contrast, ideal-noncrystal states are achieved in physical spatial dimensions through steric structure ordering towards efficient packing, and thus have fundamentally different theoretical foundations. The relationship between these states remains unclear, presenting an important problem for future investigation. Interestingly, as steric order develops towards the ideal-noncrystal state, the system loses the characteristic elastic and vibrational properties typical of amorphous solids. This observation suggests that the universal low-temperature glassy anomalies originate from the nonequilibrium and truly `disordered' structure.
We anticipate that our work may stimulate future studies.

\clearpage

\section*{Methods}
\subsection*{Basic settings}
We start with the simplest model system for amorphous solids and glass-forming liquids, consisting of polydisperse elastic particles. For proof of concept, we focus on two dimensions for ease of structure characterisations, akin to the early studies of quasicrystals \cite{levine1986quasicrystals,widom1987quasicrystal,dotera2014mosaic}. The particles interact via a finite-range repulsive potential $V(r_{ij})=\epsilon(1-r_{ij}/\sigma_{ij})^2/2$ for $r_{ij}\leq\sigma_{ij}$ and zero otherwise, where $r_{ij}$ is the distance between particles $i$ and $j$, and $\sigma_{ij}$ is the sum of their radii.
The particle diameters, $\sigma$, are chosen from a truncated power-law distribution $P(\sigma)=A\sigma^{-2}$ for $\sigma \in [\sigma_{\rm min},\sigma_{\rm max}]$, where $A$ is a normalising constant. We set $\sigma_{\rm min}/\sigma_{\rm max}\simeq0.448$, yielding a large polydispersity of $\Delta=\sqrt{\langle\sigma^2\rangle-\langle\sigma\rangle^2}/\langle\sigma\rangle \simeq 23.4\%$. This model, modified from Ref.~\onlinecite{ninarello2017models}, provides both good glass-forming ability and efficient equilibration using the swap Monte Carlo algorithm (SMC). Additionally, to achive thermodynamically favourable noncrystalline states, the particle size distribution must allow for a denser packing (at zero pressure) compared to hexagonal close packing (packing fraction $\phi=\pi/\sqrt{12}\simeq 0.907$). The upper limit of $\sigma_{\rm min}/\sigma_{\rm max}$ to meet this requirement has been mathematically derived to be around $0.65$ \cite{toth2014regular}. Our parameter setting thus falls well within the bound, and we confirm that minor variations in the parameters do not affect our results (Supplementary Information).
We fix the packing fraction at $\phi=0.92 [=\sum^{N}_{i=1}\pi\sigma_i^2/4L^2]$, slightly exceeding the upper bound of packing fraction (at zero pressure) corresponding to our choice of $\sigma_{\rm min}/\sigma_{\rm max}$ \cite{toth2014regular}. This ensures that the system remains in a solid state at zero temperature. The system is further optimized according to procedures introduced in the following to achive ideal steric order.

All particles have the same mass $m$, with the units of length, energy, temperature, and time set by the average diameter $\langle\sigma\rangle$, $\epsilon$, $\epsilon/k_B$, and  $\sqrt{m\langle\sigma\rangle^2/\epsilon}$, respectively, where $k_B$ denotes the Boltzmann constant. We equilibrate the system in the liquid state and prepare configurations by slow cooling using SMC simulations within square boxes under periodic boundary conditions. Here, the time unit is given by a full set of operations on all particles.

For the optimized ideal-noncrystal system, we also characterise the dynamics through molecular dynamics simulations in the $NVT$ ensemble with Nos{\'e}-Hoover thermostat (using LAMMPS \cite{plimpton1995fast}). From these simulations, we estimate the characteristic temperatures of the system: the onset temperature of slow glassy dynamics $T_{\rm on}$ and the hypothetical ideal glass transition temperature according to the Vogel-Fulcher-Tammann law $T_{\rm VFT}$ (Extended Data Fig.~2). The inherent states, representing zero temperature states in the nearest energy minima, are obtained using the fast inertial relaxation engine algorithm (FIRE) \cite{bitzek2006structural}. We study systems with $N=1024$ to $N=16384$ particles and confirm no significant finite-size effects (Extended Data Fig.~4). This observation is consistent with previous studies using similar forms of particle size dispersity \cite{berthier2019zero,tong2023emerging}. We mainly report results of $N=1024$, while for structural correlations (the path-integral-like correlation $C(r)$ in Fig.~\ref{fig2} and hyperuniformity in Fig.~\ref{fig4}), we present results for $N=4096$.

\subsection*{Steric order parameter}
We employ the steric order parameter $\Theta$, which quantifies the local packing capability in an order agnostic manner, to characterise the structural order of the system \cite{tong2018revealing,tong2019structural}. It has been shown that the growth of such steric order in glass-forming liquids is of thermodynamic origin \cite{tong2018revealing,tong2019structural}.
Figure~\ref{fig1}a shows a typical local configuration to illustrate the definition of $\Theta$. The neighbouring particles are identified by the radical Voronoi tessellation method \cite{okabe1992spatial,gellatly1982characterisation}. The pair $\langle ij\rangle$ of nearest neighbours and the central particle $o$ constitute a triangle, which is the basic structural unit in 2D. The imperfectness of this triangle unit is measured as the deviation of the central angle $\theta_{ij}^{(1)}$ from the perfect reference arrangement $\theta_{ij}^{(2)}$ (right panel of Fig.~\ref{fig1}a with all particles just in touch). The structural order parameter for the central particle $o$ is defined as
\begin{equation}
  \Theta_o=\sum_{\langle ij \rangle}|\theta_{ij}^{(1)}-\theta_{ij}^{(2)}|/{N_o},
\end{equation}
where $N_o$ is the number of neighbours, and the summation goes over all pairs of neighbours that are next to each other. $\Theta$ measures the deviation from sterically perfect structures: smaller $\Theta$ indicates better packing efficiency. In pursuit of ideal steric order, it is therefore desirable to tune $\Theta$ to zero.

\subsection*{Realization of ideal noncrystals}
The realization of ideal noncrystals relies on a series of constraints. First, the tendency to form crystalline (or quasicrystalline) order should be suppressed. The initial system described above is designed to fulfil this requirement. In particular, the large polydispersity frustrates crystallization, while the inverse scaling relation between the particle size and the probability distribution suppresses phase separation. In addition, guided by mathematical theorems of the packing problem, the system is set within the parameter space where the packing efficiency can exceed that of a hexagonal crystal \cite{toth2014regular}. Therefore, the hexagonal crystal can be excluded as a possible ground state in our system.
Although it is challenging to ascertain by numerical simulations that crystallisation is definitely forbidden, the free energy cost required to form complex crystals, if achievable, should be substantial.
Therefore, the ideal noncrystalline state realised in this work, even if metastable, can be highly stable and practically relevant.
We have also constrained the size difference between particles to ensure that the system can be classified as a pure phase according to Gibbs's definition, instead of a mixture of the main body (typically formed by big particles) and interstitial (small) particles.
Moreover, it is generally unknown how to construct a system through top-down design that may approach ideal steric order through thermodynamic procedures. To overcome this difficulty, we first equilibrate the system, achieving a high degree of steric order at low temperatures, and then optimise the particle sizes for ideal steric order (Fig.~\ref{fig1}b). More detailed discussions on the guiding principles to realize ideal noncrystals and the parameter dependences are provided in Supplementary Information.

Specifically, the initial system is first equilibrated at high temperature ($T=0.005$) and then cooled down to $T=10^{-6}$ in a stepwise fashion using SMC. The cooling rate is set to $dT/dt=10^{-10}$. Random displacements due to thermal fluctuations are then removed by energy minimization using FIRE. These resulting configurations are highly ordered with $\Theta=5.85\times 10^{-3}$ and serve as the starting point for the subsequent optimization procedures.
For a selected particle $o$, we adjust its size such that all its neighbours can be perfectly arranged to be just in touch, i.e., $\Theta_o=0$. This adjustment may be performed for each particle followed by energy minimization to recover mechanical equilibrium. However, to make the process efficient, we optimise the size of all particles, rescale the box size to maintain the packing fraction, and then conduct energy minimization. This optimization procedure is repeated until the average value of $\Theta$ no longer decreases and reaches $\Theta=1.55\times 10^{-3}$ (Extended Data Fig.~1). Given that the steric order is thermodynamically favourable in glass-forming liquids \cite{tong2018revealing,tong2019structural}, this procedure is expected to drive the system closer to thermodynamic equilibrium than the initial condition.
Although $\Theta$ can be further reduced by adjusting the volume fraction (e.g., using the circle packing algorithm \cite{stephenson2005introduction}), the obtained configurations are already sufficiently ordered to exhibit emergent crystal-like properties.
We note that, after the optimization procedures, the particle size distribution is close to the initial one (Fig.~\ref{fig1}c), with the polydispersity experiencing a minor change from $\Delta=23.4\%$ to $23.2\%$, and the size ratio between smallest and biggest particles changing slightly from  $\sigma_{\rm min}/\sigma_{\rm max}=0.448$ to $0.423$. These conditions strongly suggest that the pre-equilibration of the system is very weakly perturbed, ensuring the obtained ideal-noncrystal states are at, or very close to, thermodynamic equilibrium.

To confirm that the ideal-noncrystal states can indeed be approached through thermodynamic procedures, we melt the configurations obtained from optimization procedures and then perform slow cooling via SMC. We compared results from three cooling rates $dT/dt=10^{-10}$, $10^{-11}$, and $10^{-12}$ and confirmed convergence down to $T=2\times 10^{-4}$ (Extended Data Fig.~3). Therefore, we generate a large ensemble of configurations at different temperatures (corresponding to different $\Theta$) with a cooling rate $dT/dt=10^{-10}$. The corresponding inherent states are obtained using FIRE. Out-of-equilibrium data below $T=2\times 10^{-4}$ are shown with open symbols and dashed lines in the plots to distinguish them from equilibrium data.

\subsection*{Characterisation of structural correlation}
In ideal noncrystals, there is no apparent breaking of translational or rotational symmetry. Therefore, conventional methodologies, such as correlation functions of translational and orientational order used in the study of two-dimensional melting problem \cite{strandburg1988two}, cannot be applied to characterise the structural correlation in our system.
The fundamental difficulty arises from the absence of a global reference axis, against which one can define the orientations and measure the coherence of the structure.
However, upon close inspection of the ideal-noncrystal configuration, one can easily recognise the highly coherent curves formed by the outlines of particles, which extend across the whole configuration (Fig.~\ref{fig2}a). These coherent curves consistently pass through a sequence of triangles formed by three particles, which are the basic structural units in 2D.
This observation leads to a physical picture wherein the triangle units cooperate under steric constraint, extend progressively in space, and transmit interaction and information, thereby representing a unique form of structural correlation.
Inspired by this observation, we introduce the concept of `coherent paths' in ideal noncrystals, akin to lattice planes in crystals.
Although these paths are not straight in space (indicative of the absence of translational or rotational symmetry breaking), the impact of such coherent paths may build on the rigidity originating from the stability of triangle units.

The strict definition of coherent paths can be formulated as follows (inset of Fig.~\ref{fig2}a). Firstly, the neighbouring particles are determined using radical Voronoi tessellation \cite{okabe1992spatial,gellatly1982characterisation}. All triangle units formed by three neighbouring particles are identified accordingly. The radical point between two neighbouring particles is defined as the intersection of radical Voronoi and Delaunay tessellations (i.e., the intersection of inter-particle connection and the radical Voronoi tessellation). Next, we start from a selected triangle unit, choosing one of its vertices as the central particle and the rest as its neighbours. As illustrated by the inset of Fig.~\ref{fig2}a, this defines a central angle and two edges, forming the unit of a zigzag path. The connection of two radical points on the edges is identified as a unit segment of the coherent path. The coherent extension of triangles follows the zigzag form and defines the coherent path. More specifically, particles are included one by one if they are a common neighbour of the central particle and one of its neighbour. This process continues until the path returns to its initial segment (primarily due to the periodic boundary conditions), but we may also terminate the path at the boundary of the simulation box. We then proceed to select an unvisited segment (corresponding to a triangle unit with a selected central angle) and repeat the search procedure to identify a new coherent path. This is repeated until all segments are included in the coherent paths.

To characterize the structural correlation in ideal noncrystals, we introduce a path-integral-like correlation function $C(r)$. Due to the absence of a global reference axis, the structural coherence is measured along the coherent paths, by integrating the imperfectness of triangles with respect to the corresponding perfect references. As illustrated in Fig.~\ref{fig2}b, an optimal path is constructed from the original coherent path by arranging particles forming each triangle unit to be just in touch. We first focus on particles on one side of the coherent path,
similar to particles at the lattice plane in crystals, and the following procedure applies also for particles on the other side.
Starting from an arbitrary chosen pair of neighbouring particles $i$ and $j$, which defines the origin ${\bf r}={\bf r}_i=0$ and a reference direction along ${\bf r}_{ij}$, the winding angles of bonds in sequence can be determined one by one in both the original and the optimal paths. We denote the winding angle from particle $k$ to its next neighbour as $\alpha_k^{(1)}$ ($\alpha_k^{(2)}$) in the original (optimal) coherent path. We note that $\alpha_k$ can exceed the value of $\pi$ and, therefore, cannot be defined locally with respect to the reference direction of ${\bf r}_{ij}$.
Instead, the winding angle should be defined by accumulating the rotation of bond with respect to the preceding one along the coherent path $\alpha_k=\sum_{l=j}^{k}\delta \alpha_l$, with $\delta \alpha_l$ being the rotation of the bond from particle $l$ to its next neighbour with respect to the preceding bond.
Therefore, the deviation of the winding angle in the original path from that in the optimal path is caused by the accumulation of imperfectness of the constituting triangle units.
Accordingly, we can quantify the structural correlation along a coherent path
\begin{equation}\label{Eq1}
  C_p(r) = e^{6i (\alpha_k^{(1)}-\alpha_k^{(2)})} = e^{6 i\sum_{l=j}^{k}{(\delta\alpha_{l}^{(1)}-\delta\alpha_l^{(2)})}}=\prod_{l=j}^{k}{e^{6 i {(\delta\alpha_{l}^{(1)}-\delta\alpha_l^{(2)})}}},
\end{equation}
where the subscript $p$ indicates the particular path, and $r$ denotes the distance travelled from particle $i$ to $k$ along the coherent path (i.e., the summation of bond lengths). The coefficient $6$ comes from the geometric constraint that each particle has six neighbours on average in 2D.
The overall correlation function $C(r)$ is obtained by averaging over all possible paths with the same end-to-end distance along the path.
This average essentially spans all triangle units along the coherent paths, up to a distance of $r$ away. Therefore, $C(r)$ provides a comprehensive characterization of the structural correlation.
In an ideal noncrystal with perfect steric order $\Theta=0$, $C(r)$ will not decay. Conversely, in disordered liquid states, the large deviations from sterically perfect structures lead to a fast decay of $C(r)$.


Fundamentally, the path-integral-like construction of the correlation function might be linked to the inherent nature of the steric order in ideal noncrystals. Here, no particular global reference direction is preferred; instead, structural coherence is guided by steric constraint and extends in space via triangle structural units in varied directions. In the case of monodisperse systems, such structural coherence may reduce to an ordinary sixfold orientational order. However, in general, whether there exists a more fundamental description, possibly related to the Riemannian geometry with locally defined metrics~\cite{petersen2006riemannian,stephenson2005introduction}, for the steric order in ideal noncrystal states remains a topic for future studies.

\subsection*{Vibrational mode analysis}
Normal modes of vibration are fundamental to understanding the nature of solids. For inherent states in mechanical equilibrium, we obtain the normal modes by diagonalising the dynamical matrix $M={\partial}^{2}U/{\partial {\bf R}^{2}}$ using the Intel Math Kernel Library. Here $\textbf{R}=(\textbf{r}_1,\textbf{r}_2,...,\textbf{r}_N)$, and $\textbf{r}_i$ is the position of particle $i$. The vibrational density of states is calculated as $D(\omega)=\sum^{dN-d}_{j=1}\delta(\omega-\omega^j)/(dN-d)$, with $\omega^j$ being the eigenfrequency of normal mode $j$.
The participation fraction measures the degree of localization of a vibrational mode, defined as ${p}_{j}={\left(\mathop{\sum }\nolimits_{i=1}^{N}\big| {\bf{e}}_{i}^{j}{\big| }^{2}\right)}^{2}/{N\mathop{\sum }\nolimits_{i=1}^{N}\big| {\bf{e}}_{i}^{\,\,j}{\big| }^{4}}$, with ${\bf{e}}_i^j$ referring to the eigenvector of particle $i$ in mode $j$. For each condition, an ensemble of $2000$ configurations is used to ensure good statistics.

\subsection*{Characterization of elastic responses}
Distinct elastic properties between crystals and amorphous solids originate from nonaffine responses.
Therefore, we characterise the nonaffine elastic responses to probe the nature of ideal noncrystal states from a mechanical perspective.
The nonaffinity is defined as the ratio of nonaffine and affine particle displacements $\mu_{c,s}=\sum_{i}(\delta r^{c,s}_{i,{\rm NA}})^2/(\delta r^{c,s}_{i,{\rm A}})^2$ \cite{tong2015crystals}. Here, $\delta r^{c,s}_{i,{\rm NA}}$ and $\delta r^{c,s}_{i,{\rm A}}$ represent the nonaffine and affine displacement of particle $i$ under compression or shear, respectively. A strain amplitude of $\gamma=10^{-8}$ is used to ensure linear responses. For shear deformation, we adopt the Lees-Edwards boundary conditions \cite{allen2017computer}.

The scaling behaviours of elastic response close to the jamming transition provide additional information about the underlying structure \cite{liu2010jamming}. Here, we focus on the excess coordination number $\Delta z= z - z_{\rm iso}$, where $z$ is the average coordination number (excluding rattlers), and $z_{\rm iso}=2d - 2d/N$ is the isostatic value according to the Maxwell stability criterion,  with $d$ being the spatial dimension. Starting from inherent-state configurations, we decompress the system to the target pressure using the FIRE algorithm. For reference, we characterise crystals with almost perfect hexagonal geometry and extremely weak polydispersities (PC), which exhibit a new scaling relation on the edge of perfect crystalline order \cite{tong2015crystals}. The presence of such a peculiar scaling relation, different from either perfect crystals or randomly jammed packings, is a signal of ideal structural order. To ensure a fair comparison, we generate PC configurations with $N=1020$ particles and the packing fraction $\phi=0.92$. Particle size dispersity is introduced into the perfect hexagonal close packing, following a flat distribution. We carefully adjust the polydispersity to $\Delta=0.23\%$, ensuring that the steric order $\Theta=1.58\times 10^{-3}$ is close to that of ideal-noncrystal states.

\subsection*{Characterization of hyperuniformity}
To properly capture the large-scale density fluctuations in polydisperse systems, we characterise the spectral density
$\chi_{V}({\bf k})=|\sum_{j=1}^{N}{m}(|{\bf k}; R_j|)e^{-i{\bf k}\cdot {\bf r}_j}|^2/V$, where ${V}$ is the area  of the simulation box, ${m}(|{\bf k};R|) = (2\pi R/k)^{d/2}J_{d/2}(k R)$ is the Fourier transform of a disk with radius $R$, $J_\nu(x)$ is the Bessel function of order $\nu$, and $\bf{k}$ is the wave vector \cite{torquato2018hyperuniform}. The angular-averaged spectral density $\chi_{V}(k)$ is then calculated by averaging over all wave vectors of the same magnitudes. A system is considered hyperuniform if $\chi_{V}(k)\sim k^\alpha$ for $k \rightarrow 0$ and $\alpha>0$. Due to inherent imperfections, ideal hyperuniformity is not perfectly achieved, and a plateau may appear as $k\rightarrow 0$ \cite{kim2018effect}.
For comparison, we also characterise the hyperuniformity of crystals with almost perfect hexagonal geometry and extremely weak polydispersities (Extended Data Fig.~7).
In both cases, we find that the plateau value of $\chi_{V}(k)$ as $k\rightarrow 0$ exhibits a power-law scaling relation with $\Theta$.

\vspace{5mm}
\section*{Data availability}
The data that support the findings of this study are available from the corresponding authors upon request.

\section*{Code availability}
The codes that are used to generate results in the paper are available from the corresponding authors upon request.

\section*{Acknowledgments}
We thank John Russo and Ran Ni for their helpful discussions. X.F., D.X., J.Z., N.X., and H.Tong acknowledge the support of the National Natural Science Foundation of China (Grant Nos. 12274392, 12334009, and 12074355). H. Tanaka acknowledges the support of the Grant-in-Aid for Specially Promoted Research (JSPS KAKENHI Grant No. JP20H05619) from the Japan Society for the Promotion of Science (JSPS). We also thank the Supercomputing Center of the University of Science and Technology of China and the Hefei Advanced Computing Center for the computer time.

\section*{Author contributions}
H.Tong conceived the project, H.Tong, N.X., and H.Tanaka supervised the project, X.F. performed the simulations and data analysis, D.X. contributed to coding at the initial stage of the project, J.Z. contributed to the analysis of jamming scaling and hyperuniformity, all authors discussed the results, and X.F., H.Tong, H.Tanaka, and N.X. wrote the manuscript.

\section*{Competing interests}
The authors declare no competing interests.

\clearpage
\newpage


\renewcommand{\thefigure}{\arabic{figure}}
\renewcommand{\figurename}{{\bf Extended Data Fig.}}

\setcounter{figure}{0}

\begin{figure*}[htb]
  \begin{center}
    \includegraphics[width=0.6\textwidth]{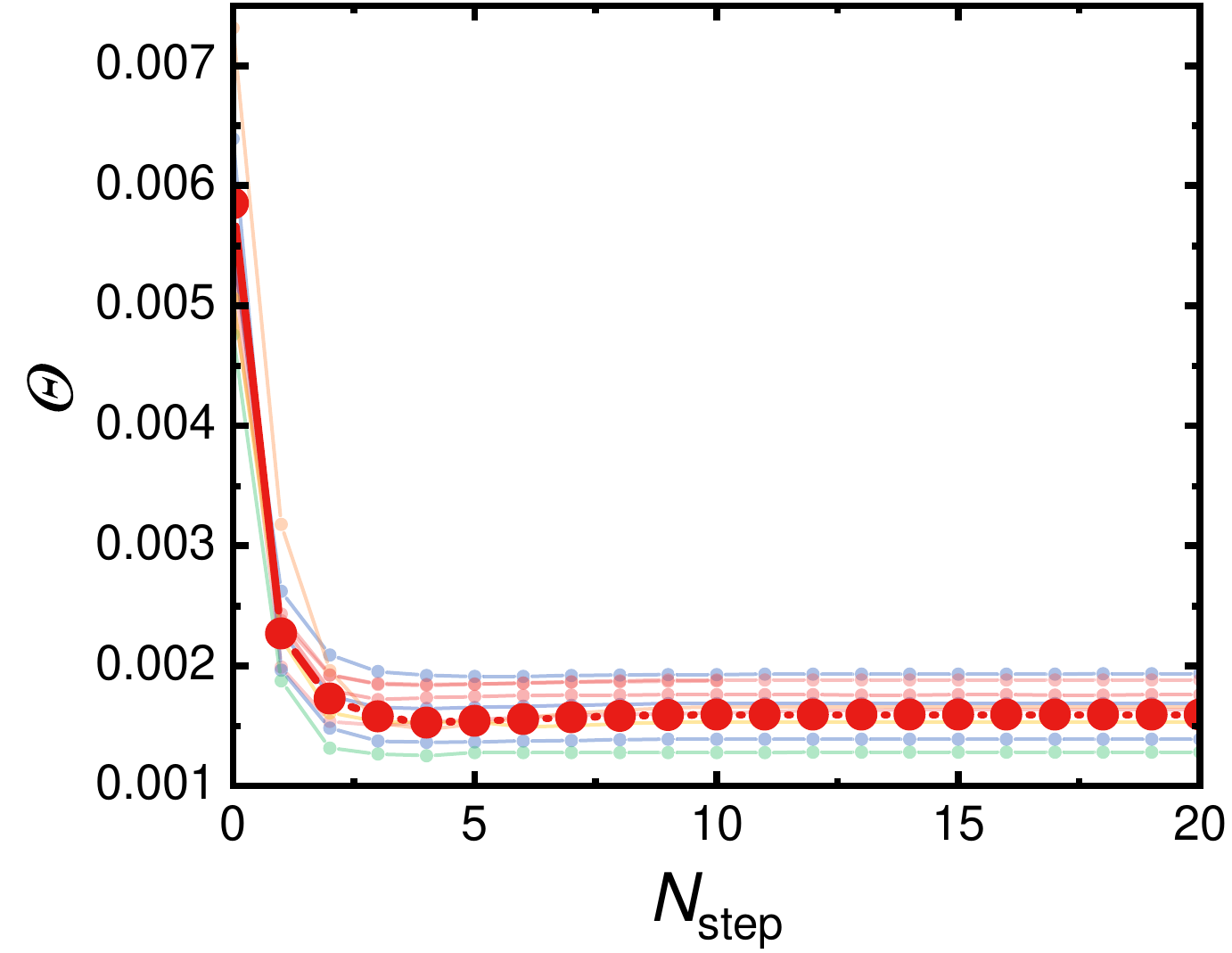}
    \caption{\textbf{Evolution of steric order $\Theta$ during the optimization procedure.} The steric order $\Theta$ as a function of iteration steps $N_{\rm step}$ for 10 independent realisations (background) and after the ensemble average (bold red circles). The convergence is rapidly achieved after around five iterations, indicating that the system is only weakly perturbed from its initial state.}
  \end{center}
    \label{figex1}
\end{figure*}
\clearpage

\begin{figure*}[htb]
  \begin{center}
    \includegraphics[width=0.98\textwidth]{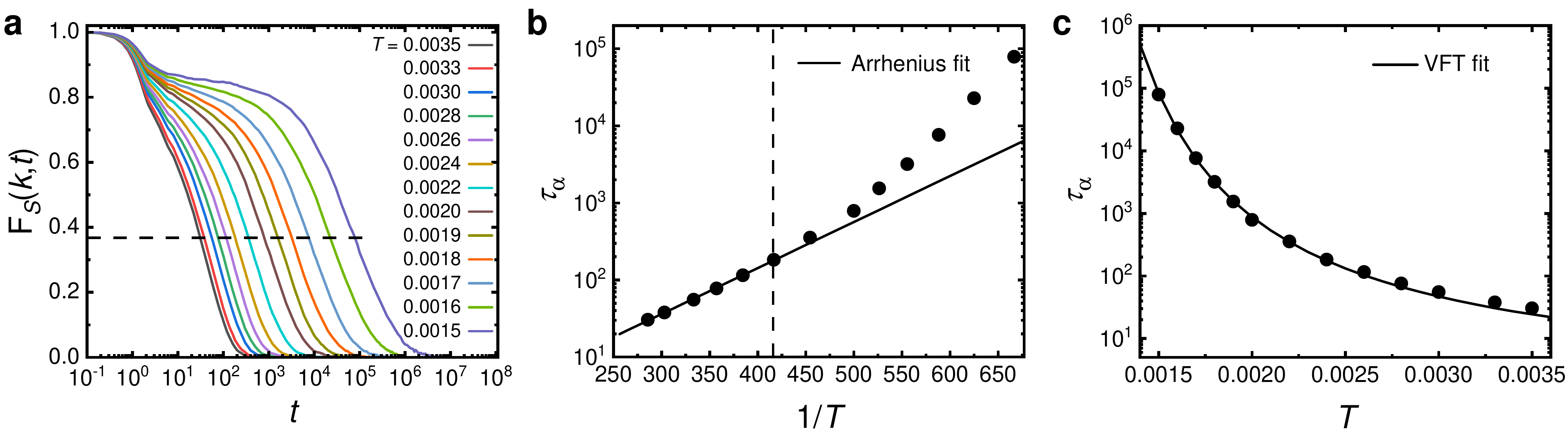}
    \caption{\textbf{Characteristic temperatures from glassy dynamics.} The dynamics of our ideal-noncrystal system is studied using molecular dynamics simulations via LAMMPS. The structure relaxation is measured by the self-intermediate scattering function $F_s(k,t)=\langle \sum_{j}{\rm exp}\left(i{\bf k}\cdot [\underline{\bf r}_j(t)-\underline{\bf r}_j(0)]\right)/N \rangle$, where $k=|{\bf k}|$ corresponds to the first peak of the static structure factor and $\langle \cdot \rangle$ denotes the time average. The relative position $\underline{\bf r}_j(t)={\bf r}_j(t)-\sum_k{\bf r}_k(t)/n_j$ is used to remove long-wavelength Mermin-Wagner fluctuations in 2D, with the summation running over all neighbours of particle $j$. The structure relaxation time $\tau_\alpha$ is defined by $F_s(k,\tau_\alpha)=e^{-1}$. \textbf{a,} Self-intermediate scattering function $F_s(k,t)$ for different temperatures. The dashed line indicates $F_s(k,t)=e^{-1}$. \textbf{b,} $\tau_\alpha$ as a function of $1/T$. The solid line shows an Arrhenius fit to the high-temperature data $\tau_\alpha \sim {\rm exp}(\Delta E/T)$. The estimated onset temperature of sluggish glassy dynamics $T_{\rm on}=2.4\times 10^{-3}$ is indicated by the dashed line.  \textbf{c,} $\tau_\alpha$ as a function of $T$. The solid line is a fit of data below $T_{\rm on}$ according to the Vogel-Fulcher-Tammann (VFT) law $\tau_\alpha \sim {\rm exp}[D T_{\rm VFT}/(T - T_{\rm VFT})]$, from which we extract the hypothesized ideal glass transition temperature $T_{\rm VFT}=9.12\times 10^{-4}$, with $D$ as a fitting parameter. $T_{\rm on}$ and $T_{\rm VFT}$ provide  two reference temperatures for our system.}
    \label{Fig4}
  \end{center}
    \label{figex2}
\end{figure*}

\begin{figure*}[htb]
  \begin{center}
    \includegraphics[width=0.85\textwidth]{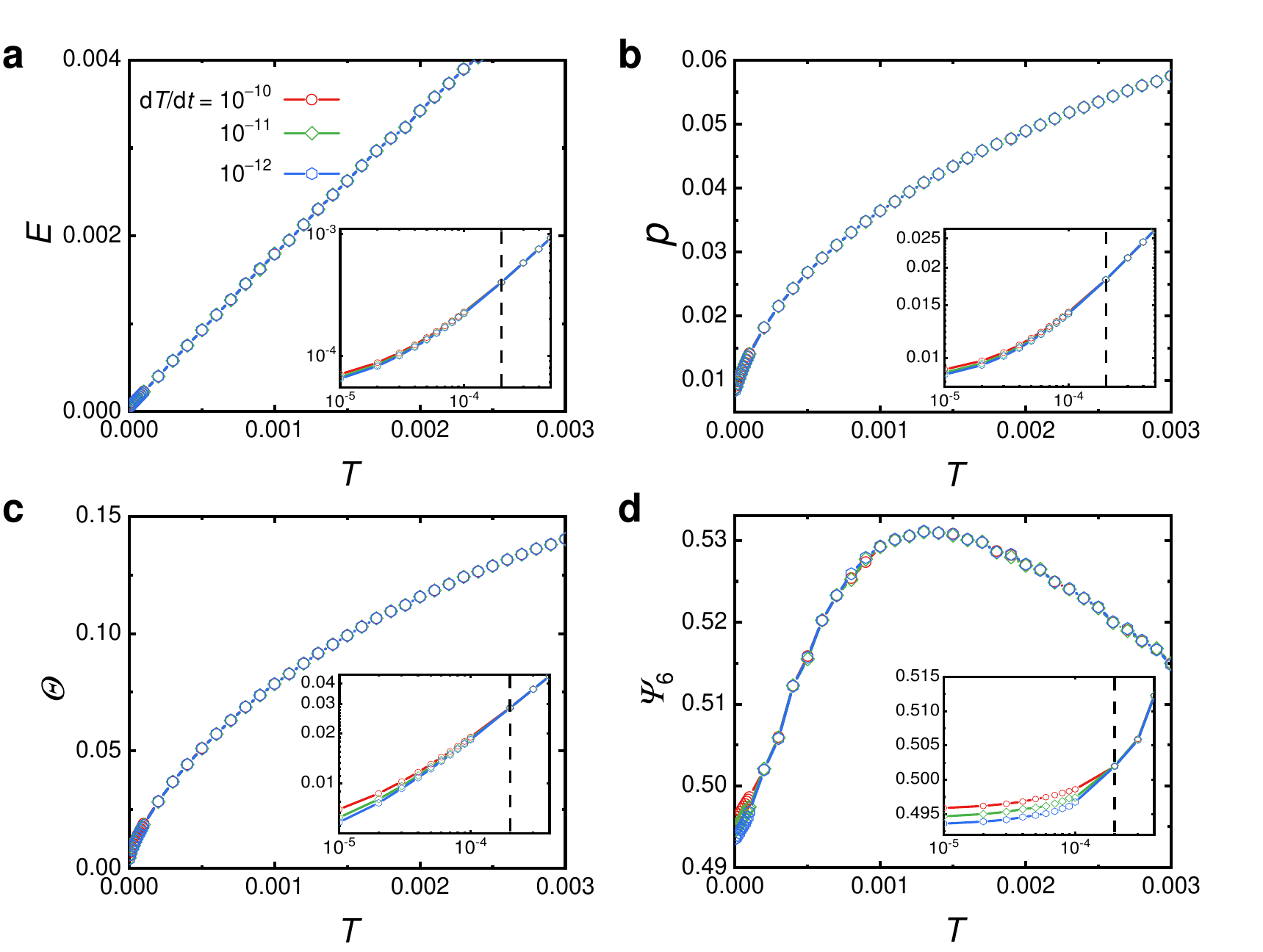}
    \caption{\textbf{Cooling rate dependence of thermodynamic quantities.} The temperature dependence of potential energy per particle $E$ (\textbf{a}), pressure $p$ (\textbf{b}), steric order parameter $\Theta$ (\textbf{c}), and hexatic bond-orientational order parameter $\Psi_6$ (\textbf{d}) for cooling rates covering three orders of magnitude. For particle $j$, $\Psi_{6}^j=\left|\sum_{k}e^{6i\theta_{jk}}/n_j \right|$, where $n_j$ is the number of nearest neighbors of particle $j$, and $\theta_{jk}$ is the angle of the bond ${\bf r}_{jk}={\bf r}_k-{\bf r}_j$ with respect to the $x$-axis. Insets provide enlarged views of the low-temperature regime. The dashed line indicates the lowest temperature, $T=2\times10^{-4}$, down to which results from different cooling rates converge, ensuring equilibration by swap Monte Carlo simulations. Importantly, in panel \textbf{d}, $\Psi_6$ shows a peak around $T=1.4\times 10^{-3}$, below which $\Psi_6$ significantly decreases with temperature, indicating that the hexatic crystalline order is thermodynamically unfavorable in our system.}
  \end{center}
    \label{figex3}
\end{figure*}
\clearpage

\begin{figure*}[htb]
  \begin{center}
    \includegraphics[width=0.85\textwidth]{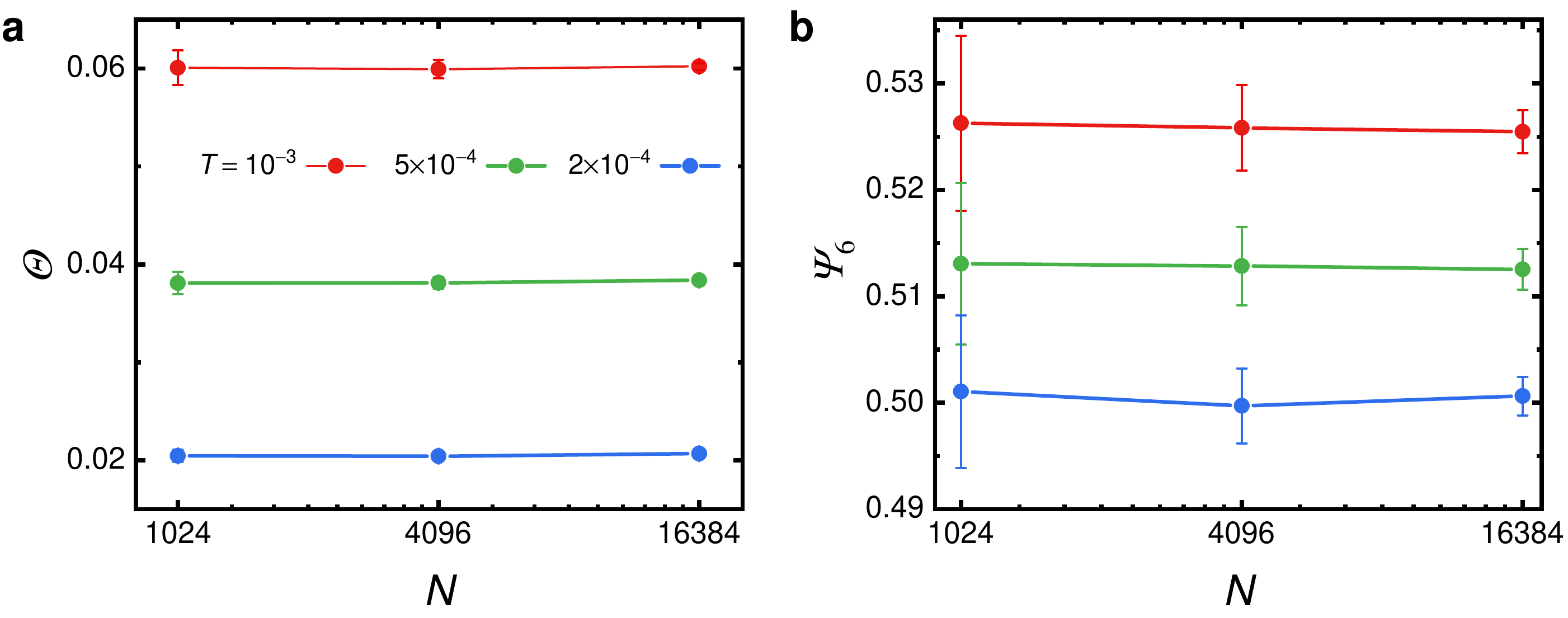}
    \caption{\textbf{System size dependence of structural order parameters.} The system size dependence of the average steric order parameter $\Theta$ (\textbf{a}) and the hexatic bond-orientational order parameter $\Psi_6$ (\textbf{b}) for three temperatures at which the system can equilibrate via swap Monte Carlo simulations. The error bars indicate the standard deviations. Both results indicate the absence of apparent finite-size effects. In particular, the value of $\Psi_6$ is quite low at low temperatures and independent of the system size, suggesting that the hexagonal crystalline order is not favoured in our system. }
  \end{center}
    \label{figex4}
\end{figure*}
\clearpage

\begin{figure*}[htb]
  \begin{center}
    \includegraphics[width=0.78\textwidth]{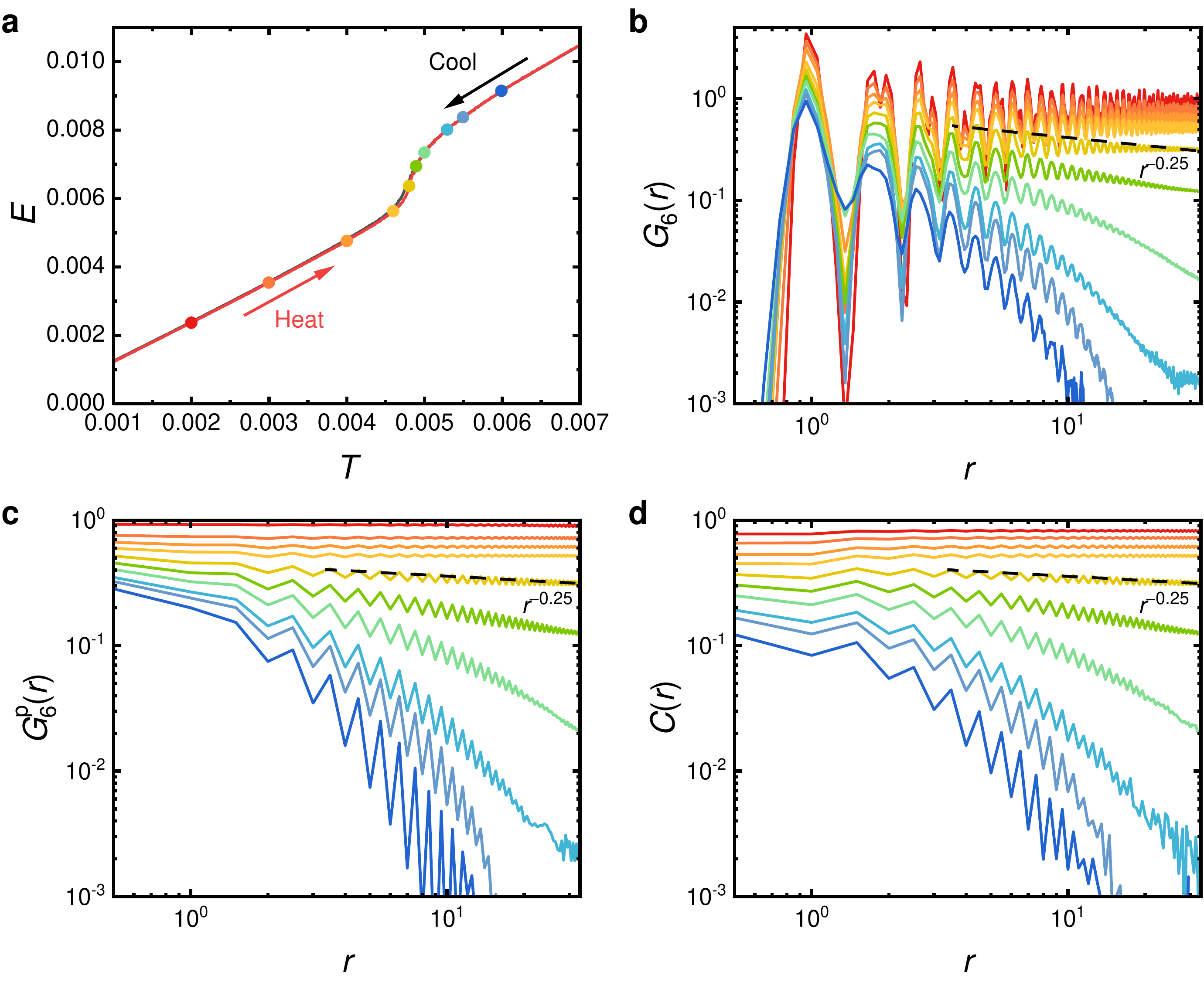}
    \caption{\textbf{Application of $C(r)$ in two-dimensional melting of monodisperse crystal.} Monodisperse crystals with the same interacting potential and packing fraction as our ideal-noncrystal system are heated and then cooled down using molecular dynamics simulations, with a constant heating (cooling) rate $dT/dt=10^{-9}$. \textbf{a,} Temperature dependence of potential energy per particle $E$ during heating (red line) and cooling (black line) processes. The evolution of $E$ during melting resembles that of ideal noncrystals, suggesting a similar physical mechanism of melting. Marginal hysteresis is observed between cooling and heating, differing from the ideal-noncrystal system. For state points indicated in \textbf{a}, the structural correlations are characterised in three different ways: The conventional correlation function of the hexatic order parameter $G_6(r)=\langle\Psi^{*}_6({\bf r})\Psi_6(0)\rangle$ (\textbf{b}), the correlation of $\Psi_6$ along the coherent path $G^{p}_6(r)$ (\textbf{c}), and the path-integral-like correlation function $C(r)$ defined in this work (\textbf{d}). The dashed line is the power-law scaling predicted by the KTHNY theory of two-dimensional melting from hexatic phase to liquids~\cite{strandburg1988two}. A close comparison confirms prefect agreement between them, validating the efficacy of our methodology. In essence, our path-integral-like scheme, based on the definition of the coherent path and the corresponding correlation function, captures the structural coherence from the steric constraint of triangle units. For monodisperse systems, the steric order reduces to the hexatic orientational order, making $C(r)$ and $G_6(r)$ give essentially the same information.}
  \end{center}
    \label{figex5}
\end{figure*}

\begin{figure*}[htb]
  \begin{center}
    \includegraphics[width=0.8\textwidth]{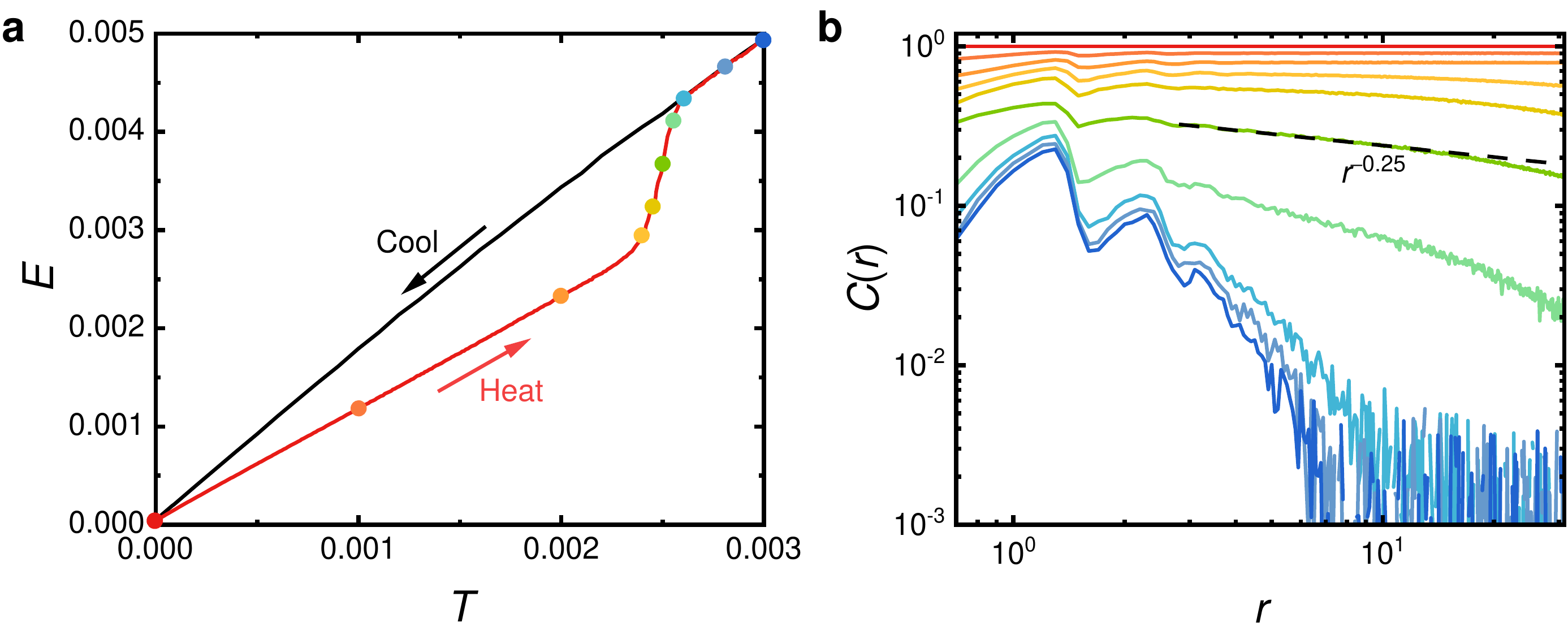}
    \caption{\textbf{Heating rate dependence of melting behavior of ideal noncrystals.} \textbf{a,} Evolution of potential energy per particle $E$ when heated from ideal-noncrystal configurations using normal MD with a heating rate $dT/dt=10^{-9}$ and then cooled down using SMC with a cooling rate $dT/dt=10^{-10}$. Compared to Fig.~\ref{fig2}c, where a heating rate of $dT/dt=10^{-10}$ is used, the steep increase of $E$ takes place at a higher temperature, but the overall behaviors are the same. This indicates a nonequilibrium nature of the melting process. \textbf{b,} The path-integral-like correlation function $C(r)$ for state points indicated in \textbf{a} (note that the temperatures are different from Fig.~\ref{fig2}d for the same colour). The dashed line indicating $C(r)\sim r^{-0.25}$ is plotted as a reference. Here, because of the faster heating rate, the system cannot be well equilibrated (even in the metastable sense) during melting. Therefore, the mixed behaviour of $C(r)$ with medium-range power-law-like correlation and long-range exponential decay at intermediate temperatures might be due to the coexistence of fluid-like and solid-like components. While melting behaviour deserves further careful investigations, the ultrastability and long-range structural correlation in ideal-noncrystal states are clear from these analyses.}
    \label{figex6}
  \end{center}
\end{figure*}

\begin{figure*}[htb]
  \begin{center}
    \includegraphics[width=0.8\textwidth]{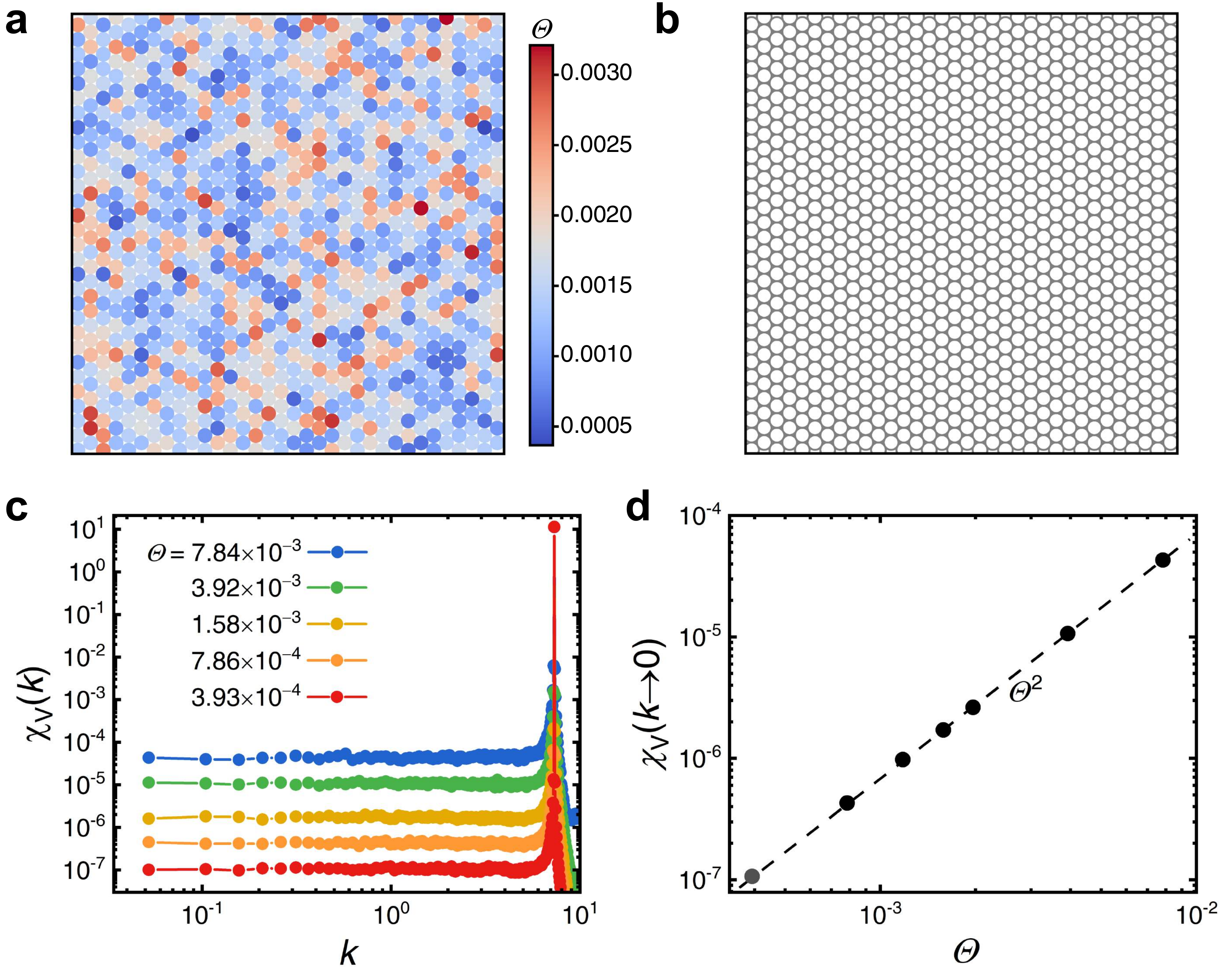}
    \caption{\textbf{The structure and hyperuniformity of weakly polydisperse crystals.} We characterize crystals with extremely weak polydispersities to demonstrate their similarity to ideal noncrystals. \textbf{a,} Visualization of a typical crystalline configuration with a polydispersity $\Delta=0.231\%$ and steric order $\Theta=1.58\times 10^{-3}$ (approximately the same as ideal noncrystals). The packing fraction is set to $\phi=0.92$ and particles are coloured according to their radii. \textbf{b,} The bare configuration corresponding to {\bf a} without colour coding, which is visually indistinguishable from a perfect hexagonal crystal. This plot gives some intuitive sense of how ordered the obtained ideal noncrystals are compared to the underlying perfect state ($\Theta=0$).
    \textbf{c,} Spectral density $\chi_V(k)$ for weakly polydisperse crystals with different degree of steric order $\Theta$. The corresponding polydispersities are $\Delta=1.155\%$, $0.577\%$ , $0.231\%$, $0.115\%$, and $0.058\%$ for decreasing $\Theta$. \textbf{d,} The plateau value of spectral density when approaching the low-$k$ limit $\chi_V(k\rightarrow 0)$ as a function of steric order $\Theta$. The dashed line is a power-law extrapolation fitting of the data $\chi_V(k\rightarrow 0)\sim \Theta^{2}$. Similar results were obtained in a recent study of defected crystals, where imperfections are introduced by point vacancies and interstitials \cite{kim2018effect}.}
  \end{center}
    \label{figex7}
\end{figure*}


\begin{thebibliography}{10}
\expandafter\ifx\csname url\endcsname\relax
  \def\url#1{\texttt{#1}}\fi
\expandafter\ifx\csname urlprefix\endcsname\relax\def\urlprefix{URL }\fi
\providecommand{\bibinfo}[2]{#2}
\providecommand{\eprint}[2][]{\url{#2}}

\bibitem{ashcroft1976solid}
\bibinfo{author}{Ashcroft, N.~W.} \& \bibinfo{author}{Mermin, N.~D.}
\newblock \emph{\bibinfo{title}{Solid State Physics}}
  (\bibinfo{publisher}{Thomson Brooks/Cole, Belmont, MA},
  \bibinfo{year}{1976}).

\bibitem{phillips1981amorphous}
\bibinfo{editor}{Phillips, W.~A.} (ed.) \emph{\bibinfo{title}{Amorphous Solids:
  Low-Temperature Properties}} (\bibinfo{publisher}{Springer},
  \bibinfo{year}{1981}).

\bibitem{shechtman1984metallic}
\bibinfo{author}{Shechtman, D.}, \bibinfo{author}{Blech, I.},
  \bibinfo{author}{Gratias, D.} \& \bibinfo{author}{Cahn, J.~W.}
\newblock \bibinfo{title}{Metallic phase with long-range orientational order
  and no translational symmetry}.
\newblock \emph{\bibinfo{journal}{Phys. Rev. Lett.}}
  \textbf{\bibinfo{volume}{53}}, \bibinfo{pages}{1951} (\bibinfo{year}{1984}).

\bibitem{levine1984quasicrystals}
\bibinfo{author}{Levine, D.} \& \bibinfo{author}{Steinhardt, P.~J.}
\newblock \bibinfo{title}{Quasicrystals: {A} new class of ordered structures}.
\newblock \emph{\bibinfo{journal}{Phys. Rev. Lett.}}
  \textbf{\bibinfo{volume}{53}}, \bibinfo{pages}{2477} (\bibinfo{year}{1984}).

\bibitem{levine1986quasicrystals}
\bibinfo{author}{Levine, D.} \& \bibinfo{author}{Steinhardt, P.~J.}
\newblock \bibinfo{title}{Quasicrystals. {I}. definition and structure}.
\newblock \emph{\bibinfo{journal}{Phys. Rev. B}} \textbf{\bibinfo{volume}{34}},
  \bibinfo{pages}{596} (\bibinfo{year}{1986}).

\bibitem{divincenzo1991quasicrystals}
\bibinfo{author}{Divincenzo, D.} \& \bibinfo{author}{Steinhardt, P.~J.}
\newblock \bibinfo{title}{Quasicrystals: The state of the art}
  (\bibinfo{year}{1991}).

\bibitem{pauling1985apparent}
\bibinfo{author}{Pauling, L.}
\newblock \bibinfo{title}{Apparent icosahedral symmetry is due to directed
  multiple twinning of cubic crystals}.
\newblock \emph{\bibinfo{journal}{Nature}} \textbf{\bibinfo{volume}{317}},
  \bibinfo{pages}{512--514} (\bibinfo{year}{1985}).

\bibitem{steinhardt2019second}
\bibinfo{author}{Steinhardt, P.}
\newblock \emph{\bibinfo{title}{The second kind of impossible: The
  extraordinary quest for a new form of matter}} (\bibinfo{publisher}{Simon and
  Schuster}, \bibinfo{year}{2019}).

\bibitem{international1992report}
\bibinfo{title}{Report of the {E}xecutive {C}ommittee for 1991}.
\newblock \emph{\bibinfo{journal}{Acta Cryst.}} \textbf{\bibinfo{volume}{A48}},
  \bibinfo{pages}{922--946} (\bibinfo{year}{1992}).

\bibitem{zeng2004supramolecular}
\bibinfo{author}{Zeng, X.} \emph{et~al.}
\newblock \bibinfo{title}{Supramolecular dendritic liquid quasicrystals}.
\newblock \emph{\bibinfo{journal}{Nature}} \textbf{\bibinfo{volume}{428}},
  \bibinfo{pages}{157--160} (\bibinfo{year}{2004}).

\bibitem{iacovella2011self}
\bibinfo{author}{Iacovella, C.~R.}, \bibinfo{author}{Keys, A.~S.} \&
  \bibinfo{author}{Glotzer, S.~C.}
\newblock \bibinfo{title}{Self-assembly of soft-matter quasicrystals and their
  approximants}.
\newblock \emph{\bibinfo{journal}{Proceedings of the National Academy of
  Sciences}} \textbf{\bibinfo{volume}{108}}, \bibinfo{pages}{20935--20940}
  (\bibinfo{year}{2011}).

\bibitem{plati2024quasi}
\bibinfo{author}{Plati, A.} \emph{et~al.}
\newblock \bibinfo{title}{Quasi-crystalline order in vibrating granular
  matter}.
\newblock \emph{\bibinfo{journal}{Nat. Phys.}} \bibinfo{pages}{1--7}
  (\bibinfo{year}{2024}).

\bibitem{steurer2018quasicrystals}
\bibinfo{author}{Steurer, W.}
\newblock \bibinfo{title}{Quasicrystals: {W}hat do we know? {W}hat do we want
  to know? {W}hat can we know?}
\newblock \emph{\bibinfo{journal}{Acta Cryst.}} \textbf{\bibinfo{volume}{74}},
  \bibinfo{pages}{1--11} (\bibinfo{year}{2018}).

\bibitem{steinhardt1983bond}
\bibinfo{author}{Steinhardt, P.~J.}, \bibinfo{author}{Nelson, D.~R.} \&
  \bibinfo{author}{Ronchetti, M.}
\newblock \bibinfo{title}{Bond-orientational order in liquids and glasses}.
\newblock \emph{\bibinfo{journal}{Phys. Rev. B}} \textbf{\bibinfo{volume}{28}},
  \bibinfo{pages}{784--805} (\bibinfo{year}{1983}).

\bibitem{alexander1998amorphous}
\bibinfo{author}{Alexander, S.}
\newblock \bibinfo{title}{Amorphous solids: their structure, lattice dynamics
  and elasticity}.
\newblock \emph{\bibinfo{journal}{Phys. Rep.}} \textbf{\bibinfo{volume}{296}},
  \bibinfo{pages}{65--236} (\bibinfo{year}{1998}).

\bibitem{tanaka2012bond}
\bibinfo{author}{Tanaka, H.}
\newblock \bibinfo{title}{Bond orientational order in liquids: Towards a
  unified description of water-like anomalies, liquid-liquid transition, glass
  transition, and crystallization}.
\newblock \emph{\bibinfo{journal}{Eur. Phys. J. E}}
  \textbf{\bibinfo{volume}{35}}, \bibinfo{pages}{113} (\bibinfo{year}{2012}).

\bibitem{tanaka2019revealing}
\bibinfo{author}{Tanaka, H.}, \bibinfo{author}{Tong, H.}, \bibinfo{author}{Shi,
  R.} \& \bibinfo{author}{Russo, J.}
\newblock \bibinfo{title}{Revealing key structural features hidden in liquids
  and glasses}.
\newblock \emph{\bibinfo{journal}{Nat. Rev. Phys.}} \bibinfo{pages}{1}
  (\bibinfo{year}{2019}).

\bibitem{steinhardt1981icosahedral}
\bibinfo{author}{Steinhardt, P.~J.}, \bibinfo{author}{Nelson, D.~R.} \&
  \bibinfo{author}{Ronchetti, M.}
\newblock \bibinfo{title}{Icosahedral bond orientational order in supercooled
  liquids}.
\newblock \emph{\bibinfo{journal}{Phys. Rev. Lett.}}
  \textbf{\bibinfo{volume}{47}}, \bibinfo{pages}{1297} (\bibinfo{year}{1981}).

\bibitem{nelson1983order}
\bibinfo{author}{Nelson, D.~R.}
\newblock \bibinfo{title}{Order, frustration, and defects in liquids and
  glasses}.
\newblock \emph{\bibinfo{journal}{Phys. Rev. B}} \textbf{\bibinfo{volume}{28}},
  \bibinfo{pages}{5515} (\bibinfo{year}{1983}).

\bibitem{penrose1974role}
\bibinfo{author}{Penrose, R.}
\newblock \bibinfo{title}{The role of aesthetics in pure and applied
  mathematical research}.
\newblock \emph{\bibinfo{journal}{Bull. Inst. Math. Appl.}}
  \textbf{\bibinfo{volume}{10}}, \bibinfo{pages}{266--271}
  (\bibinfo{year}{1974}).

\bibitem{berthier2011theoretical}
\bibinfo{author}{Berthier, L.} \& \bibinfo{author}{Biroli, G.}
\newblock \bibinfo{title}{Theoretical perspective on the glass transition and
  amorphous materials}.
\newblock \emph{\bibinfo{journal}{Rev. Mod. Phys.}}
  \textbf{\bibinfo{volume}{83}}, \bibinfo{pages}{587--645}
  (\bibinfo{year}{2011}).

\bibitem{frank1958complex}
\bibinfo{author}{Frank, F.~C.} \& \bibinfo{author}{Kasper, J.~S.}
\newblock \bibinfo{title}{Complex alloy structures regarded as sphere packings.
  {I}. {D}efinitions and basic principles}.
\newblock \emph{\bibinfo{journal}{Acta Cryst.}} \textbf{\bibinfo{volume}{11}},
  \bibinfo{pages}{184--190} (\bibinfo{year}{1958}).

\bibitem{bernal1959geometrical}
\bibinfo{author}{Bernal, J.~D.}
\newblock \bibinfo{title}{A geometrical approach to the structure of liquids}.
\newblock \emph{\bibinfo{journal}{Nature}} \textbf{\bibinfo{volume}{183}},
  \bibinfo{pages}{141--147} (\bibinfo{year}{1959}).

\bibitem{torquato2018perspective}
\bibinfo{author}{Torquato, S.}
\newblock \bibinfo{title}{Perspective: Basic understanding of condensed phases
  of matter via packing models}.
\newblock \emph{\bibinfo{journal}{J. Chem. Phys.}}
  \textbf{\bibinfo{volume}{149}} (\bibinfo{year}{2018}).

\bibitem{xing2024origin}
\bibinfo{author}{Xing, Y.} \emph{et~al.}
\newblock \bibinfo{title}{Origin of the critical state in sheared granular
  materials}.
\newblock \emph{\bibinfo{journal}{Nat. Phys.}} \bibinfo{pages}{1--7}
  (\bibinfo{year}{2024}).

\bibitem{tong2018revealing}
\bibinfo{author}{Tong, H.} \& \bibinfo{author}{Tanaka, H.}
\newblock \bibinfo{title}{Revealing hidden structural order controlling both
  fast and slow glassy dynamics in supercooled liquids}.
\newblock \emph{\bibinfo{journal}{Phys. Rev. X}} \textbf{\bibinfo{volume}{8}},
  \bibinfo{pages}{011041} (\bibinfo{year}{2018}).

\bibitem{tong2019structural}
\bibinfo{author}{Tong, H.} \& \bibinfo{author}{Tanaka, H.}
\newblock \bibinfo{title}{Structural order as a genuine control parameter of
  dynamics in simple glass formers}.
\newblock \emph{\bibinfo{journal}{Nat. Commun.}} \textbf{\bibinfo{volume}{10}},
  \bibinfo{pages}{4899} (\bibinfo{year}{2019}).

\bibitem{ninarello2017models}
\bibinfo{author}{Ninarello, A.}, \bibinfo{author}{Berthier, L.} \&
  \bibinfo{author}{Coslovich, D.}
\newblock \bibinfo{title}{Models and algorithms for the next generation of
  glass transition studies}.
\newblock \emph{\bibinfo{journal}{Phy. Rev. X}} \textbf{\bibinfo{volume}{7}},
  \bibinfo{pages}{021039} (\bibinfo{year}{2017}).

\bibitem{gibbs1878equilibrium}
\bibinfo{author}{Gibbs, J.~W.}
\newblock \bibinfo{title}{On the equilibrium of heterogeneous substances}.
\newblock \emph{\bibinfo{journal}{Trans. Conn. Acad. Arts Sci.}}
  \textbf{\bibinfo{volume}{3}}, \bibinfo{pages}{108–248}
  (\bibinfo{year}{1878}).

\bibitem{brito2018theory}
\bibinfo{author}{Brito, C.}, \bibinfo{author}{Lerner, E.} \&
  \bibinfo{author}{Wyart, M.}
\newblock \bibinfo{title}{Theory for swap acceleration near the glass and
  jamming transitions for continuously polydisperse particles}.
\newblock \emph{\bibinfo{journal}{Phys. Rev. X}} \textbf{\bibinfo{volume}{8}},
  \bibinfo{pages}{031050} (\bibinfo{year}{2018}).

\bibitem{kim2022structural}
\bibinfo{author}{Kim, S.} \& \bibinfo{author}{Hilgenfeldt, S.}
\newblock \bibinfo{title}{Structural measures as guides to ultrastable states
  in overjammed packings}.
\newblock \emph{\bibinfo{journal}{Phys. Rev. Lett.}}
  \textbf{\bibinfo{volume}{129}}, \bibinfo{pages}{168001}
  (\bibinfo{year}{2022}).

\bibitem{kim2024dense}
\bibinfo{author}{Kim, S.} \& \bibinfo{author}{Hilgenfeldt, S.}
\newblock \bibinfo{title}{Exceptionally dense and resilient polydisperse disk
  packings}.
\newblock \emph{\bibinfo{journal}{arXiv:2402.08390}}  (\bibinfo{year}{2024}).

\bibitem{corwin2024ideal}
\bibinfo{author}{Bolton-Lum, V.}, \bibinfo{author}{Dennis, R.~C.},
  \bibinfo{author}{Morse, P.} \& \bibinfo{author}{Corwin, E.}
\newblock \bibinfo{title}{The ideal glass and the ideal disk packing in two
  dimensions}.
\newblock \emph{\bibinfo{journal}{arXiv:2404.07492}}  (\bibinfo{year}{2024}).

\bibitem{halperin1978theory}
\bibinfo{author}{Halperin, B.~I.} \& \bibinfo{author}{Nelson, D.~R.}
\newblock \bibinfo{title}{Theory of two-dimensional melting}.
\newblock \emph{\bibinfo{journal}{Phys. Rev. Lett.}}
  \textbf{\bibinfo{volume}{41}}, \bibinfo{pages}{121} (\bibinfo{year}{1978}).

\bibitem{strandburg1988two}
\bibinfo{author}{Strandburg, K.~J.}
\newblock \bibinfo{title}{Two-dimensional melting}.
\newblock \emph{\bibinfo{journal}{Rev. Mod. Phys.}}
  \textbf{\bibinfo{volume}{60}}, \bibinfo{pages}{161} (\bibinfo{year}{1988}).

\bibitem{kosterlitz1973ordering}
\bibinfo{author}{Kosterlitz, J.~M.} \& \bibinfo{author}{Thouless, D.~J.}
\newblock \bibinfo{title}{Ordering, metastability and phase transitions in
  two-dimensional systems}.
\newblock \emph{\bibinfo{journal}{J. Phys. C: Solid State Phys.}}
  \textbf{\bibinfo{volume}{6}}, \bibinfo{pages}{1181} (\bibinfo{year}{1973}).

\bibitem{nelson1979dislocation}
\bibinfo{author}{Nelson, D.~R.} \& \bibinfo{author}{Halperin, B.~I.}
\newblock \bibinfo{title}{Dislocation-mediated melting in two dimensions}.
\newblock \emph{\bibinfo{journal}{Phys. Rev. B}} \textbf{\bibinfo{volume}{19}},
  \bibinfo{pages}{2457} (\bibinfo{year}{1979}).

\bibitem{young1979melting}
\bibinfo{author}{Young, A.~P.}
\newblock \bibinfo{title}{Melting and the vector coulomb gas in two
  dimensions}.
\newblock \emph{\bibinfo{journal}{Phys. Rev. B}} \textbf{\bibinfo{volume}{19}},
  \bibinfo{pages}{1855} (\bibinfo{year}{1979}).

\bibitem{lerner2016statistics}
\bibinfo{author}{Lerner, E.}, \bibinfo{author}{D{\"u}ring, G.} \&
  \bibinfo{author}{Bouchbinder, E.}
\newblock \bibinfo{title}{Statistics and properties of low-frequency
  vibrational modes in structural glasses}.
\newblock \emph{\bibinfo{journal}{Phys. Rev. Lett.}}
  \textbf{\bibinfo{volume}{117}}, \bibinfo{pages}{035501}
  (\bibinfo{year}{2016}).

\bibitem{richard2020universality}
\bibinfo{author}{Richard, D.} \emph{et~al.}
\newblock \bibinfo{title}{Universality of the nonphononic vibrational spectrum
  across different classes of computer glasses}.
\newblock \emph{\bibinfo{journal}{Phys. Rev. Lett.}}
  \textbf{\bibinfo{volume}{125}}, \bibinfo{pages}{085502}
  (\bibinfo{year}{2020}).

\bibitem{tong2015crystals}
\bibinfo{author}{Tong, H.}, \bibinfo{author}{Tan, P.} \& \bibinfo{author}{Xu,
  N.}
\newblock \bibinfo{title}{From crystals to disordered crystals: A hidden
  order-disorder transition}.
\newblock \emph{\bibinfo{journal}{Sci. Rep.}} \textbf{\bibinfo{volume}{5}},
  \bibinfo{pages}{15378} (\bibinfo{year}{2015}).

\bibitem{maloney2006amorphous}
\bibinfo{author}{Maloney, C.~E.} \& \bibinfo{author}{Lemaitre, A.}
\newblock \bibinfo{title}{Amorphous systems in athermal, quasistatic shear}.
\newblock \emph{\bibinfo{journal}{Phys. Rev. E}} \textbf{\bibinfo{volume}{74}},
  \bibinfo{pages}{016118} (\bibinfo{year}{2006}).

\bibitem{van2009jamming}
\bibinfo{author}{van Hecke, M.}
\newblock \bibinfo{title}{Jamming of soft particles: geometry, mechanics,
  scaling and isostaticity}.
\newblock \emph{\bibinfo{journal}{J. Phys. Condens. Matter}}
  \textbf{\bibinfo{volume}{22}}, \bibinfo{pages}{033101}
  (\bibinfo{year}{2010}).

\bibitem{liu2010jamming}
\bibinfo{author}{Liu, A.~J.} \& \bibinfo{author}{Nagel, S.~R.}
\newblock \bibinfo{title}{The jamming transition and the marginally jammed
  solid}.
\newblock \emph{\bibinfo{journal}{Annu. Rev. Condens. Matter Phys.}}
  \textbf{\bibinfo{volume}{1}}, \bibinfo{pages}{347--369}
  (\bibinfo{year}{2010}).

\bibitem{torquato2018hyperuniform}
\bibinfo{author}{Torquato, S.}
\newblock \bibinfo{title}{Hyperuniform states of matter}.
\newblock \emph{\bibinfo{journal}{Phys. Rep.}} \textbf{\bibinfo{volume}{745}},
  \bibinfo{pages}{1--95} (\bibinfo{year}{2018}).

\bibitem{zachary2011hyperuniform}
\bibinfo{author}{Zachary, C.~E.}, \bibinfo{author}{Jiao, Y.} \&
  \bibinfo{author}{Torquato, S.}
\newblock \bibinfo{title}{Hyperuniform long-range correlations are a signature
  of disordered jammed hard-particle packings}.
\newblock \emph{\bibinfo{journal}{Phys. Rev. Lett.}}
  \textbf{\bibinfo{volume}{106}}, \bibinfo{pages}{178001}
  (\bibinfo{year}{2011}).

\bibitem{lei2019nonequilibrium}
\bibinfo{author}{Lei, Q.-L.}, \bibinfo{author}{Ciamarra, M.~P.} \&
  \bibinfo{author}{Ni, R.}
\newblock \bibinfo{title}{Nonequilibrium strongly hyperuniform fluids of circle
  active particles with large local density fluctuations}.
\newblock \emph{\bibinfo{journal}{Sci. Adv.}} \textbf{\bibinfo{volume}{5}},
  \bibinfo{pages}{eaau7423} (\bibinfo{year}{2019}).

\bibitem{chen2024emergent}
\bibinfo{author}{Chen, J.} \emph{et~al.}
\newblock \bibinfo{title}{Emergent chirality and hyperuniformity in an active
  mixture with nonreciprocal interactions}.
\newblock \emph{\bibinfo{journal}{Phys. Rev. Lett.}}
  \textbf{\bibinfo{volume}{132}}, \bibinfo{pages}{118301}
  (\bibinfo{year}{2024}).

\bibitem{torquato2015ensemble}
\bibinfo{author}{Torquato, S.}, \bibinfo{author}{Zhang, G.} \&
  \bibinfo{author}{Stillinger, F.~H.}
\newblock \bibinfo{title}{Ensemble theory for stealthy hyperuniform disordered
  ground states}.
\newblock \emph{\bibinfo{journal}{Phys. Rev. X}} \textbf{\bibinfo{volume}{5}},
  \bibinfo{pages}{021020} (\bibinfo{year}{2015}).

\bibitem{kim2018effect}
\bibinfo{author}{Kim, J.} \& \bibinfo{author}{Torquato, S.}
\newblock \bibinfo{title}{Effect of imperfections on the hyperuniformity of
  many-body systems}.
\newblock \emph{\bibinfo{journal}{Phys. Rev B}} \textbf{\bibinfo{volume}{97}},
  \bibinfo{pages}{054105} (\bibinfo{year}{2018}).

\bibitem{yunker2014physics}
\bibinfo{author}{Yunker, P.~J.} \emph{et~al.}
\newblock \bibinfo{title}{Physics in ordered and disordered colloidal matter
  composed of poly ({N}-isopropylacrylamide) microgel particles}.
\newblock \emph{\bibinfo{journal}{Rep. Prog. Phys.}}
  \textbf{\bibinfo{volume}{77}}, \bibinfo{pages}{056601}
  (\bibinfo{year}{2014}).

\bibitem{corker2019}
\bibinfo{author}{Corker, A.}, \bibinfo{author}{Ng, H. C.-H.},
  \bibinfo{author}{Poole, R.~J.} \& \bibinfo{author}{Garc{\'i}a-Tu{\~n}{\'o}n,
  E.}
\newblock \bibinfo{title}{3{D} printing with 2{D} colloids: designing rheology
  protocols to predict ‘printability’ of soft-materials}.
\newblock \emph{\bibinfo{journal}{Soft Matter}} \textbf{\bibinfo{volume}{15}},
  \bibinfo{pages}{1444--1456} (\bibinfo{year}{2019}).

\bibitem{widom1987quasicrystal}
\bibinfo{author}{Widom, M.}, \bibinfo{author}{Strandburg, K.~J.} \&
  \bibinfo{author}{Swendsen, R.~H.}
\newblock \bibinfo{title}{Quasicrystal equilibrium state}.
\newblock \emph{\bibinfo{journal}{Phys. Rev. Lett.}}
  \textbf{\bibinfo{volume}{58}}, \bibinfo{pages}{706} (\bibinfo{year}{1987}).

\bibitem{dotera2014mosaic}
\bibinfo{author}{Dotera, T.}, \bibinfo{author}{Oshiro, T.} \&
  \bibinfo{author}{Ziherl, P.}
\newblock \bibinfo{title}{Mosaic two-lengthscale quasicrystals}.
\newblock \emph{\bibinfo{journal}{Nature}} \textbf{\bibinfo{volume}{506}},
  \bibinfo{pages}{208--211} (\bibinfo{year}{2014}).

\bibitem{toth2014regular}
\bibinfo{author}{T{\'o}th, L.~F.}
\newblock \emph{\bibinfo{title}{Regular figures}}
  (\bibinfo{publisher}{Elsevier}, \bibinfo{year}{2014}).

\bibitem{plimpton1995fast}
\bibinfo{author}{Plimpton, S.}
\newblock \bibinfo{title}{Fast parallel algorithms for short-range molecular
  dynamics}.
\newblock \emph{\bibinfo{journal}{J. Comput. Phys}}
  \textbf{\bibinfo{volume}{117}}, \bibinfo{pages}{1--19}
  (\bibinfo{year}{1995}).

\bibitem{bitzek2006structural}
\bibinfo{author}{Bitzek, E.}, \bibinfo{author}{Koskinen, P.},
  \bibinfo{author}{G{\"a}hler, F.}, \bibinfo{author}{Moseler, M.} \&
  \bibinfo{author}{Gumbsch, P.}
\newblock \bibinfo{title}{Structural relaxation made simple}.
\newblock \emph{\bibinfo{journal}{Phys. Rev. Lett.}}
  \textbf{\bibinfo{volume}{97}}, \bibinfo{pages}{170201}
  (\bibinfo{year}{2006}).

\bibitem{berthier2019zero}
\bibinfo{author}{Berthier, L.}, \bibinfo{author}{Charbonneau, P.},
  \bibinfo{author}{Ninarello, A.}, \bibinfo{author}{Ozawa, M.} \&
  \bibinfo{author}{Yaida, S.}
\newblock \bibinfo{title}{Zero-temperature glass transition in two dimensions}.
\newblock \emph{\bibinfo{journal}{Nat. Commun.}} \textbf{\bibinfo{volume}{10}},
  \bibinfo{pages}{4875} (\bibinfo{year}{2019}).

\bibitem{tong2023emerging}
\bibinfo{author}{Tong, H.} \& \bibinfo{author}{Tanaka, H.}
\newblock \bibinfo{title}{Emerging exotic compositional order on approaching
  low-temperature equilibrium glasses}.
\newblock \emph{\bibinfo{journal}{Nat. Commun.}} \textbf{\bibinfo{volume}{14}},
  \bibinfo{pages}{4614} (\bibinfo{year}{2023}).

\bibitem{okabe1992spatial}
\bibinfo{author}{Okabe, A.}
\newblock \emph{\bibinfo{title}{Spatial Tessellations — Concepts and
  Applications of Voronoi Diagrams}} (\bibinfo{publisher}{John Wiley \& Sons,
  Inc.}, \bibinfo{year}{1992}).

\bibitem{gellatly1982characterisation}
\bibinfo{author}{Gellatly, B.~J.} \& \bibinfo{author}{Finney, J.~L.}
\newblock \bibinfo{title}{Characterisation of models of multicomponent
  amorphous metals: the radical alternative to the voronoi polyhedron}.
\newblock \emph{\bibinfo{journal}{J. Non-Cryst. Solids}}
  \textbf{\bibinfo{volume}{50}}, \bibinfo{pages}{313--329}
  (\bibinfo{year}{1982}).

\bibitem{stephenson2005introduction}
\bibinfo{author}{Stephenson, K.}
\newblock \emph{\bibinfo{title}{Introduction to circle packing: The theory of
  discrete analytic functions}} (\bibinfo{publisher}{Cambridge University
  Press}, \bibinfo{year}{2005}).

\bibitem{petersen2006riemannian}
\bibinfo{author}{Petersen, P.}
\newblock \emph{\bibinfo{title}{Riemannian geometry}}, vol.
  \bibinfo{volume}{171} (\bibinfo{publisher}{Springer}, \bibinfo{year}{2006}).

\bibitem{allen2017computer}
\bibinfo{author}{Allen, M.~P.} \& \bibinfo{author}{Tildesley, D.~J.}
\newblock \emph{\bibinfo{title}{Computer simulation of liquids}}
  (\bibinfo{publisher}{Oxford university press}, \bibinfo{year}{2017}).

\end{thebibliography}
\end{document}